\documentclass[11pt,hyper,a4paper]{article}
\usepackage{jheppub}
\usepackage{amsmath,amssymb,amsfonts,amscd,bm, amsthm, slashed, mathrsfs, bm}
\usepackage{graphicx, wrapfig}
\usepackage{multirow}
\usepackage{verbatim}
\usepackage[title]{appendix}
\usepackage{fancybox}
\usepackage[utf8]{inputenc}

\usepackage{url}
\usepackage{float}

\usepackage[normalem]{ulem}
\usepackage{graphicx}

\usepackage{color}
\usepackage[usenames,dvipsnames,svgnames,table]{xcolor}
\definecolor{darkblue}{cmyk}{0.9,0.9,0,0}
\hypersetup{colorlinks=false, linkcolor=darkblue,linkcolor=darkblue, citecolor=darkblue}

\usepackage{multirow}
\usepackage{verbatim}

\usepackage{bbm}
\usepackage{comment}
\usepackage{enumitem}
\usepackage{float}
\usepackage{empheq}
\usepackage[usenames,dvipsnames,svgnames,table]{xcolor}
\usepackage{enumerate}

\title{BPS Invariants for 3-Manifolds at Rational Level $K$}

\author{Hee-Joong Chung}

\affiliation{Yau Mathematical Sciences Center, Tsinghua University, Haidian District, Beijing 100084, China}

\abstract{We consider the Witten-Reshetikhin-Turaev invariants or Chern-Simons partition functions at or around roots of unity $q=e^{\frac{2\pi i}{K}}$ with a rational level $K=\frac{r}{s}$ where $r$ and $s$ are coprime integers.
From the exact expression for the $G=SU(2)$ Witten-Reshetikhin-Turaev invariants of the Seifert manifolds at a rational level obtained by Lawrence and Rozansky,
we provide an expected form of the structure of the Witten-Reshetikhin-Turaev invariants in terms of the homological blocks at a rational level.
Also, we discuss the asymptotic expansion of knot invariants around roots of unity where we take a limit different from the limit in the standard volume conjecture.
}

\begin{document}

\maketitle

%%%%%%%%%%%%%%%%%%%%%%%%%%%%%%%%%%%%%%%%%%%%%%%%%%%%%%%%%%%%%%%%%%%%%%%%%%%%%%%%%%%%%%
%%%%%%%%%%%%%%%%%%%%%%%%%%%%%%%%%%%%%%%%%%%%%%%%%%%%%%%%%%%%%%%%%%%%%%%%%%%%%%%%%%%%%%
%%%%%%%%%%%%%%%%%%%%%%%%%%%%%%%%%%%%%%%%%%%%%%%%%%%%%%%%%%%%%%%%%%%%%%%%%%%%%%%%%%%%%%
%%%%%%%%%%%%%%%%%%%%%%%%%%%%%%%%%%%%%%%%%%%%%%%%%%%%%%%%%%%%%%%%%%%%%%%%%%%%%%%%%%%%%%
%%%%%%%%%%%%%%%%%%%%%%%%%%%%%%%%%%%%%%%%%%%%%%%%%%%%%%%%%%%%%%%%%%%%%%%%%%%%%%%%%%%%%%

\section{Introduction}

It is known that the Chern-Simons (CS) partition function agrees with the Reshetikhin-Turaev invariant, and they are often called the WRT (Witten-Reshetikhin-Turaev) invariant \cite{Witten-Jones, RT}.
Differently from the usual Chern-Simons theory where the level $K$ is taken to be an integer, $K$ can be a rational number in the Reshetikhin-Turaev construction.
The Chern-Simons theory with an integer level $K$ and its analytic continuation has been studied actively in the literature, so it would be interesting to consider the case with $K$ being a rational number and its analytic continuation from a rational $K$.	\\

In this paper, we discuss the properties and the behavior of the WRT invariant or the CS partition function at or around roots of unity for closed 3-manifolds and knot complements in $S^3$.

In section \ref{hom-block}, we consider the WRT invariant at other roots of unity, \textit{i.e.} $q=e^{2\pi i /K}$ with a rational $K$, in terms of homological blocks.
It was conjectured that the WRT invariant for the standard root of unity $q=e^{2\pi i \frac{1}{K}}$ where $K$ is an integer can be expressed in terms of the so called homological block, which is a $q$-series invariant with integer powers and integer coefficients \cite{Gukov-Putrov-Vafa,Gukov-Pei-Putrov-Vafa}.\footnote{For recent developments of homological blocks, we refer to \cite{Gukov-Marino-Putrov,Chun:2017dbf,Cheng:2018vpl,Chung-wrt,Gukov:2019mnk}}
From the integral formula of the $SU(2)$ WRT invariants for a certain infinite family of Seifert manifolds at other roots of unity $q=e^{\frac{2\pi i}{K}}$ with $K=\frac{r}{s}$ where $r$ and $s$ are coprime integers \cite{Lawrence-Rozansky}, we express the WRT invariant in terms of homological blocks.
We find that the general structure is similar to the case of the integer $K$ and also provide explicit expressions for some examples.

In section \ref{asymp-exp}, we consider an asymptotic expansion of knot invariants around general roots of unity.
In the context of the volume conjecture, the asymptotic expansion in the limit of the level $K \rightarrow \infty$ has been discussed in the literature.
This asymptotic expansion is the expansion around $q \rightarrow 1$ where the deviation from 1 is $e^{\frac{2\pi i}{K}}$.
The limit around general roots of unity which we consider for the asymptotic expansion is different from the limit taken in the usual volume conjecture.
We consider the asymptotic expansion around a given root of unity $e^{\frac{2\pi i}{K}}$ with a rational level $K = \frac{r}{s}$ where the deviation from $e^{2\pi i \frac{s}{r}}$ is given by another expansion parameter.
This was discussed in \cite{Chung-rou}, also in \cite{Gukov-Pei-rou, Dimofte:2015kkp, Garoufalidis:2018qds} in a similar but a slightly different setup or in a different context.
We calculate the leading order of the asymptotic expansion of superpolynomials for some knots around roots of unity.

In section \ref{discussion}, we summarize and discuss some future directions.	\\

\noindent \textit{Note added:} While preparing the manuscript, we found that \cite{Kucharski:2019fgh} appeared, which overlaps with parts of section \ref{hom-block} of this paper.

%%%%%%%%%%%%%%%%%%%%%%%%%%%%%%%%%%%%%%%%%%%%%%%%%%%%%%%%%%%%%%%%%%%%%%%%%%%%%%%%%%%%%%
%%%%%%%%%%%%%%%%%%%%%%%%%%%%%%%%%%%%%%%%%%%%%%%%%%%%%%%%%%%%%%%%%%%%%%%%%%%%%%%%%%%%%%
%%%%%%%%%%%%%%%%%%%%%%%%%%%%%%%%%%%%%%%%%%%%%%%%%%%%%%%%%%%%%%%%%%%%%%%%%%%%%%%%%%%%%%
%%%%%%%%%%%%%%%%%%%%%%%%%%%%%%%%%%%%%%%%%%%%%%%%%%%%%%%%%%%%%%%%%%%%%%%%%%%%%%%%%%%%%%
%%%%%%%%%%%%%%%%%%%%%%%%%%%%%%%%%%%%%%%%%%%%%%%%%%%%%%%%%%%%%%%%%%%%%%%%%%%%%%%%%%%%%%

\section{WRT invariants and homological blocks for $G=SU(2)$ at other roots of unity}
\label{hom-block}

In \cite{Lawrence-Rozansky}, in addition to the case of the standard root of unity $q=e^{\frac{2\pi i}{K}}$ where $K \in \mathbb{Z}$, the WRT invariant of a certain infinite family of Seifert manifolds at other roots of unity was also obtained.
As in the case of the standard root of unity, we would like to express the WRT invariant at other roots of unity in terms of the homological blocks.

%%%%%%%%%%%%%%%%%%%%%%%%%%%%%%%%%%%%%%%%%%%%%%%%%%%%%%%%%%%%%%%%%%%%%%%%%%%%%%%%%%%%%%
%%%%%%%%%%%%%%%%%%%%%%%%%%%%%%%%%%%%%%%%%%%%%%%%%%%%%%%%%%%%%%%%%%%%%%%%%%%%%%%%%%%%%%
%%%%%%%%%%%%%%%%%%%%%%%%%%%%%%%%%%%%%%%%%%%%%%%%%%%%%%%%%%%%%%%%%%%%%%%%%%%%%%%%%%%%%%

\subsection{WRT invariant at other roots of unity}

For Seifert manifolds $X(P_1/Q_1, \ldots, P_F/Q_F)$ with conditions that $P_j$'s and $Q_j$'s are coprime for each $j$, $j=1,\ldots, F$, and $P_j$'s are pairwise coprime, the finite sum expression for the WRT invariant in \cite{Lawrence-Rozansky} is given by
\begin{align}
Z_{K}(M_3) = \frac{B}{K} e^{\frac{\pi i}{2K} \phi_F} \sum_{\substack{\beta = -PK \\ K \nmid \beta}}^{PK} e^{-\frac{\pi i}{2K} \frac{H}{P} \beta^2} 
\frac{\prod_{j=1}^{F} e^{\frac{\pi i \beta}{K P_j} } - e^{-\frac{\pi i \beta}{K P_j} }}{(e^{\frac{\pi i \beta}{K}}-e^{\frac{\pi i \beta}{K}})^{F-2}}	\label{wrt-finite-sum}
\end{align}
where $H := P \sum_{i=1}^3 \frac{Q_i}{P_i} = \pm |\text{Tor} \, H_1(M, \mathbb{Z}) |$ and $P = \prod_{j=1}^F P_j$.
Also, 
\begin{align}
B	&=	-\frac{\text{sign}P}{4 \sqrt{|P|}} e^{\frac{3}{4}\pi i \, \text{sign} \, \left( \frac{H}{P} \right)}	\,	,	\\
\phi_F	&=	3 \, \text{sign}\left( \frac{H}{P} \right) + \sum_{j=1}^{F} \left( 12 s(Q_j, P_j) - \frac{Q_j}{P_j} \right)		\,	,
\end{align}
where $s(Q,P)$ is the Dedekind sum
\begin{align}
s(Q,P) = \frac{1}{4P} \sum_{l=1}^{P-1} \cot \Big( \frac{\pi l}{P} \Big) \cot \Big( \frac{\pi Q l}{P} \Big)	\,	
\end{align}
for $P>0$, which satisfies $s(-Q,P)=-s(Q,P)$.
Here, we use the physics normalization
\begin{align}
Z_{K}(S^1 \times S^2) = 1	\,	,	\qquad	Z_{K}(S^3) = \sqrt{\frac{2}{K}} \sin \Big(\frac{\pi}{K}\Big)	\,	.
\end{align}
With $x=e^{2\pi i \frac{1}{4KP}}$, \eqref{wrt-finite-sum} can be written as
\begin{align}
Z_{K}(M_3) = \frac{B}{K} x^{P\phi_F} \sum_{\substack{\beta = -PK \\ K \nmid \beta}}^{PK} x^{- H \beta^2} 
\frac{\prod_{j=1}^{F} x^{\frac{\beta P}{P_j} } - x^{-\frac{\beta P}{P_j} }}{(x^{2P\beta}-e^{-2P\beta})^{F-2}}	\,	.	\label{wrt-finite-sum-x}
\end{align}
Therefore, the expression \eqref{wrt-finite-sum-x} includes the part that is in $\mathbb{Q}[x]$.
By considering the Galois action on $x$, $x$ is replaced with another primitive root of unity, $e^{2\pi i \frac{s}{4KP}}$, where $s$ is coprime to $4KP$.
Due to this change, an overall factor appears from the factor $B$, and we don't consider it in this paper.

Up to such an overall factor, the integral expression for the WRT invariant at other roots of unity $e^{\frac{2\pi i}{K}}$ with $K=\frac{r}{s}$ for the Seifert manifolds $X(P_1/Q_1, \ldots, P_F/Q_F)$ is given by
\begin{align}
Z_{K=\frac{r}{s}}(M_3) \simeq \frac{B}{2 \pi i s} e^{-\frac{2\pi i}{K} \frac{\phi_3}{4}} \Bigg( \sum_{t=0}^{Hs-1} \int_C dy f(y) e^{-2\pi i t y} - 2\pi i \sum_{m=0}^{2Ps-1} \text{Res } \bigg( \frac{f(y)}{1-e^{-2\pi i y}}, y=m K \bigg)	\Bigg)	\label{wrt-rational}
\end{align}
where
\begin{align}
f(y) = e^{- \frac{2\pi i}{K} \frac{H}{4P} y^2} \frac{\prod_{j=1}^F (e^{\frac{2\pi i}{K} \frac{y}{2P_j}} - e^{-\frac{2\pi i}{K} \frac{y}{2P_j}})}{ e^{\frac{2\pi i}{K} \frac{y}{2}} - e^{-\frac{2\pi i}{K} \frac{y}{2}}}
\end{align}
and $C$ is a contour that passes from $(-1+i)\infty$, 0, to $(1-i)\infty$ for $\frac{P}{H}>0$ and a clockwise rotation of it for $\frac{P}{H}<0$.

%%%%%%%%%%%%%%%%%%%%%%%%%%%%%%%%%%%%%%%%%%%%%%%%%%%%%%%%%%%%%%%%%%%%%%%%%%%%%%%%%%%%%%
%%%%%%%%%%%%%%%%%%%%%%%%%%%%%%%%%%%%%%%%%%%%%%%%%%%%%%%%%%%%%%%%%%%%%%%%%%%%%%%%%%%%%%
%%%%%%%%%%%%%%%%%%%%%%%%%%%%%%%%%%%%%%%%%%%%%%%%%%%%%%%%%%%%%%%%%%%%%%%%%%%%%%%%%%%%%%

\subsection{WRT invariant at other roots of unity in terms of homological blocks}

As we have done for the case of the standard root of unity, we would like to analytically continue the level from $K=\frac{r}{s}$, and express the WRT invariant in terms of the homological blocks.
We closely follow the calculation in \cite{Chung-wrt}.
For simplicity, we consider the case $F=3$ and $\sum_{j=1}^3 \frac{1}{P_j} <1$.\footnote{For the case of $F=3$, $\sum_{j=1}^3 \frac{1}{P_j}>1$ can only happen when $(P_1,P_2,P_3)=(2,3,5)$ and in this case an additional term proportional to $2 q^{\frac{1}{120}}$ arises.
}

The Gaussian integral part of \eqref{wrt-rational} can be written as
\begin{align}
\frac{B}{2\pi i s} e^{-\phi_3/4} \sum_{t=0}^{Hs-1} \int_{\Gamma_t} dy \, e^{-\frac{1}{2\pi i} \frac{r}{s} \frac{H}{P} \big( y + 2\pi i \frac{P}{H} t \big)^2 + 2\pi i \frac{P}{H} \frac{r}{s} t^2} 
\frac{\prod_{j=1}^3 (e^{ \frac{y}{P_j}} - e^{- \frac{y}{P_j}})}{ e^{y} - e^{-y}}	\label{gaussian}
\end{align}
where the integration cycle $\Gamma_t$ is chosen in such a way that for each $t$ the integrand is convergent on both ends of infinity.
That is, when $r/s >0 $ and $P/H>0$, $\Gamma_0$ is a line from $-(1+i)\infty$ to $(1+i)\infty$ through the origin. 
$\Gamma_t$ is parallel to $\Gamma_0$ and passes through a stationary phase point $y=-2\pi i \frac{P}{H}t$.
When $P/H<0$, the contour is given by a clockwise rotation of $\Gamma_0$ by $\frac{\pi}{2}$ and similarly for $\Gamma_t$.

Assuming that $\text{Re }y >0$ and $P>0$, the rational function of sine hyperbolic functions can be expanded in terms of a periodic function $\chi_{2P}(n)$, 
\begin{align}
\frac{\prod_{j=1}^3 (e^{ \frac{y}{P_j}} - e^{- \frac{y}{P_j}})}{ e^{y} - e^{-y}} = \sum_{n=0}^\infty \chi_{2P} (n) e^{-\frac{n}{P}y}	\label{expansion}
\end{align}
and $\chi_{2P}(n)$ is expressed in terms of another periodic function $\psi_{2P}^{(l)}(n)$
\begin{align}
\chi_{2P}(n) = \sum_{c=0}^3 \psi^{(R_c)}_{2P}(n)
\end{align}
where
\begin{align}
\psi^{(l)}_{2P}(n) = \begin{cases} \pm 1	&	\text{if } n \equiv \pm l	\quad	\text{mod } 2P	\\	0	&	\text{otherwise} \end{cases}
\end{align}
and $R_0=P(1-(1/{P_1} + 1/{P_2} + 1/{P_3}))$, $R_1 = P(1-(1/{P_1} - 1/{P_2} - 1/{P_3}))$, $R_2=P(1-(-1/{P_1} + 1/{P_2} - 1/{P_3}))$, and $R_3=P(1-(-1/{P_1} - 1/{P_2} + 1/{P_3}))$.
Given $r$ and $s$, we consider the analytic continuation of $K$ from $\frac{r}{s}$.
We also assume $H>0$ and $\text{Im } K <0$ for convergence.
We take a contour $\gamma$ as in the case of standard root of unity, \textit{i.e.} a line parallel to the imaginary axis of the $y$-plane that passes through $\text{Re} \, y >0$.
Then the integral above with the contour $\gamma$ gives
\begin{align}
\frac{B}{2 i} q^{-\phi_3/4} \bigg( \frac{2i}{r/s} \frac{P}{H} \bigg)^{1/2} \sum_{t=0}^{Hs-1} e^{2\pi i K \frac{P}{H} t^2} \sum_{n=0}^\infty \chi_{2P}(n) e^{2\pi i \frac{t}{H} n} q^{\frac{n^2}{4HP}}
\label{wrt-rs-gen}
\end{align}
where we chose an overall factor to have $\frac{1}{\sqrt{r/s}}$ by multiplying $s$.

As in the case of the standard root of unity, we expect that the expression that we obtained from the Gaussian integral part of the WRT invariant with analytic continuation from $K=\frac{r}{s}$ knows the contributions from all flat connections.
Also, the calculation can be done for other ranges of $P$, $H$, and $\text{Im}\frac{r}{s}$, which we refer to section 2.7 in \cite{Chung-wrt} for detailed explanation.

%%%%%%%%%%%%%%%%%%%%%%%%%%%%%%%%%%%%%%%%%%%%%%%%%%%%%%%%%%%%%%%%%%%%%%%%%%%%%%%%%%%%%%
%%%%%%%%%%%%%%%%%%%%%%%%%%%%%%%%%%%%%%%%%%%%%%%%%%%%%%%%%%%%%%%%%%%%%%%%%%%%%%%%%%%%%%
%%%%%%%%%%%%%%%%%%%%%%%%%%%%%%%%%%%%%%%%%%%%%%%%%%%%%%%%%%%%%%%%%%%%%%%%%%%%%%%%%%%%%%

\subsection{The case $H=1$}

We consider the case $H=1$ first.
In this case, we have
\begin{align}
Z_{K=\frac{r}{s}}(M_3) = \frac{B}{2 i} q^{-\phi_3/4} \bigg( \frac{2iP}{r/s}\bigg)^{1/2}  \sum_{t=0}^{s-1} e^{2\pi i \frac{r}{s} P t^2} \sum_{n=0}^\infty \chi_{2P}(n) q^{\frac{n^2}{4P}} \bigg|_{q \searrow e^{2\pi i \frac{s}{r}}}	\,	.
\end{align}
Here, $\sum_{t=0}^{s-1} e^{2\pi i \frac{r}{s} P t^2}$ is the quadratic Gauss sum $g(m;s)$ which is defined as
\begin{align}
g(m;s) = \sum_{n=0}^{s-1} e^{2\pi i m n^2 / s}	\,	.
\end{align}
Therefore, in this case, we obtain
\begin{align}
Z_{K=\frac{r}{s}}(M_3) = \frac{B}{2 i} q^{-\phi_3/4} \bigg( \frac{2iP}{r/s}\bigg)^{1/2} g(Pr;s) \sum_{i=1}^{3} \widetilde{\Psi}_{2P}^{(R_i)}(q)  \bigg|_{q \searrow e^{2\pi i \frac{s}{r}}}	\,		\label{wrt-rs-h1}
\end{align}
where $\widetilde{\Psi}_{P}^{(l)}(q)$ is a false theta function
\begin{align}
\widetilde{\Psi}_{P}^{(l)}(q) = \sum_{n=0}^\infty \psi^{(l)}_{2P}(n) e^{2\pi i \frac{1}{4KP} n^2} = \sum_{n=0}^\infty \psi^{(l)}_{2P}(n) q^{\frac{n^2}{4P}}	\,	,
\end{align}
which is the Eichler integral of the modular form $\Psi^{(l)}_P(q) := \sum_{n=0}^{\infty} n \psi_{2P}^{(l)}(n) q^{\frac{n^2}{4P}}$ of half-integer weight $3/2$ \cite{Lawrence-Zagier}.
In the limit $q \searrow e^{2\pi i \frac{s}{r}}$, $\widetilde{\Psi}_{P}^{(l)} (q)$ becomes
\begin{align}
\widetilde{\Psi}_{P}^{(l)} (e^{2\pi i \frac{s}{r}}) = \sum_{n=0}^{rP} \Big( 1- \frac{n}{rP} \Big) \psi^{(l)}_{2P}(n) e^{2\pi i \frac{1}{4P} \frac{s}{r}n^2}	\label{ftheta-f-sum}
\end{align}
where we refer to \cite{Lawrence-Zagier, Zagier-identity} (see also \cite{Hikami-Kirilov, Hikami-torus}) for derivation.
$s$ is coprime to $4rP$, so $s$ is odd here.

We also note that when $s=1$, \eqref{wrt-rs-h1} becomes the WRT invariant with the standard root of unity.
When $H=1$, there is only one homological block, which is well defined in the region $|q|<1$.
Therefore, as in the case of the standard root of unity, we may expect that the WRT invariant at $q = e^{2\pi i \frac{s}{r}}$ would be obtained from the limit $q \rightarrow e^{2\pi i \frac{s}{r}}$ of a homological block and above calculation indicates that it is so up to an overall factor.

%%%%%%%%%%%%%%%%%%%%%%%%%%%%%%%%%%%%%%%%%%%%%%%%%%%%%%%%%%%%%%%%%%%%%%%%%%%%%%%%%%%%%%
%%%%%%%%%%%%%%%%%%%%%%%%%%%%%%%%%%%%%%%%%%%%%%%%%%%%%%%%%%%%%%%%%%%%%%%%%%%%%%%%%%%%%%
%%%%%%%%%%%%%%%%%%%%%%%%%%%%%%%%%%%%%%%%%%%%%%%%%%%%%%%%%%%%%%%%%%%%%%%%%%%%%%%%%%%%%%

\subsection{The case $H\geq 2$}
\label{ssec:h2}

When $H \geq 2$, we decompose $\psi^{(l)}_{2P}(n)$ in terms of $\psi^{(l)}_{2HP}(n)$
\begin{align}
\psi_{2P}^{(l)}(n) = \sum_{h=0}^{\lceil \frac{H}{2}-1 \rceil} \psi_{2HP}^{(2hP+l)}(n) - \sum_{h=0}^{\lfloor \frac{H}{2}-1 \rfloor} \psi_{2HP}^{(2(h+1)P+l)}(n)
\end{align}
where the floor and the ceiling function are given by
\begin{align}
\lfloor x \rfloor 	&= \text{max} \{ m \in \mathbb{Z} \, | \, m \leq x \}	\,	,	\\
\lceil x \rceil 	&= \text{min} \{ m \in \mathbb{Z} \, | \, m \geq x \}	\,	.
\end{align}

As done in the case of the standard root of unity \cite{Chung-wrt}, $e^{2\pi i \frac{t}{H}n}$ can be taken out of the summation in \eqref{wrt-rs-gen}. 
We repeat it for completeness.
We take a representative in \eqref{wrt-rs-gen}
\begin{align}
\sum_{t=1}^{Hs-1} e^{2\pi i K \frac{P}{H}t^2} \sum_{n=0}^{\infty} e^{2\pi i \frac{t}{H}n} \psi_{2HP}^{(l)}(n) q^{\frac{1}{4HP} n^2}	\,	,	\label{3f-preresult}
\end{align}
which is nonzero when $n=2HPm+l$ and $2HPm'-l$ with $m, m' \in \mathbb{Z}_{\geq 0}$.
The $n=2HPm+l$ part of \eqref{3f-preresult} is 
\begin{align}
\sum_{t=1}^{Hs-1} e^{2\pi i K \frac{P}{H}t^2} \sum_{\substack{n=2HPm+l \\ m \in \mathbb{Z}_{\geq 0}}}^{\infty} e^{2\pi i \frac{t}{H}l} \psi_{2HP}^{(l)}(n) q^{\frac{1}{4HP} n^2}	
= \sum_{t=1}^{Hs-1} e^{2\pi i K \frac{P}{H}t^2} e^{2\pi i \frac{t}{H}l} \sum_{\substack{n=2HPm+l \\ m \in \mathbb{Z}_{\geq 0}}}^{\infty} \psi_{2HP}^{(l)}(n) q^{\frac{1}{4HP} n^2}	\label{3f-pre-pos}	
\end{align}
where $e^{2\pi i \frac{t}{H}(2HPm+l)} = e^{2\pi i \frac{t}{H}l}$ is used.
The $n=2HPm-l$ part of \eqref{3f-preresult} is given by
\begin{align}
\sum_{t=1}^{Hs-1} e^{2\pi i K \frac{P}{H}t^2} \sum_{\substack{n=2HPm-l \\ m \in \mathbb{Z}_{\geq 0}}}^{\infty} e^{-2\pi i \frac{t}{H}l} \psi_{2HP}^{(l)}(n) q^{\frac{1}{4HP} n^2}	
= \sum_{t=1}^{Hs-1} e^{2\pi i K \frac{P}{H}t^2} e^{-2\pi i \frac{t}{H}l} \sum_{\substack{n=2HPm-l \\ m \in \mathbb{Z}_{\geq 0}}}^{\infty} \psi_{2HP}^{(l)}(n) q^{\frac{1}{4HP} n^2}	\label{3f-pre-neg}	\,	,
\end{align}
which can be written as
\begin{align}
\sum_{t'=1}^{Hs-1} e^{2\pi i r s PH-4\pi i rPt'+2\pi i \frac{r}{s} \frac{P}{H}t'^2} e^{2\pi i \frac{t'}{H}l} \sum_{\substack{n=2HPm-l \\ m \in \mathbb{Z}_{\geq 0}}}^{\infty} \psi_{2HP}^{(l)}(n) q^{\frac{1}{4HP}}	\label{3f-pre-neg2}	\,	
\end{align}
with $t'=Hs-t$.
Since we take a limit to $K=\frac{r}{s}$ with $r$ and $s$ being coprime integers, $e^{2\pi i r s PH-4\pi i rPt'+2\pi i \frac{r}{s} \frac{P}{H}t'^2}$ becomes $e^{2\pi i \frac{r}{s} \frac{P}{H}t'^2}$.
Thus, \eqref{3f-preresult} can be expressed as
\begin{align}
\sum_{t=1}^{Hs-1} e^{2\pi i K \frac{P}{H}t^2} e^{2\pi i \frac{t}{H}l} \sum_{n=0}^{\infty} \psi_{2HP}^{(l)}(n) q^{\frac{1}{4HP}n^2}	\,	.
\end{align}
Therefore, \eqref{wrt-rs-gen} can be written as
\begin{align}
\begin{split}
Z_{K=\frac{r}{s}}(M_3)
\simeq&
\Bigg[
\sum_{t=0}^{Hs-1} e^{2\pi i K \frac{P}{H} t^2} \sum_{m=0}^{3} \Big(  \sum_{h=0}^{\lceil \frac{H}{2}-1 \rceil} e^{2\pi i \frac{t}{H} (2hP+R_m)} \widetilde{\Psi}^{(2hP+R_m)}_{HP}(q)	\\
&\hspace{35mm} - \sum_{h=0}^{\lfloor \frac{H}{2}-1 \rfloor} e^{2\pi i \frac{t}{H} (2(h+1)P-R_m)} \widetilde{\Psi}^{(2(h+1)P-R_m)}_{HP}(q) \Big) 		
\Bigg]  \Bigg|_{q \searrow e^{2\pi i \frac{s}{r}}}	\,	.
\end{split}	\label{wrt-rs-hblock}
\end{align}

The expression \eqref{wrt-rs-hblock} can be further organized.
We first consider the case that $H$ is odd.
The case for even $H$ can be done similarly.
We take $t=0,1, \ldots, Hs-1$ as $t=Hv+u$ where $u=0,1, \ldots, H-1$ and $v=0,1, \ldots, s-1$,
\begin{align}
\hspace{-10mm}\sum_{v=0}^{s-1} \sum_{u=0}^{H-1} e^{2\pi i \frac{r}{s} \frac{P}{H} (Hv+u)^2} 
\sum_{m=0}^{3} \Big(  \sum_{h=0}^{\frac{H-1}{2}} e^{2\pi i \frac{u}{H} (2hP+R_m)} \widetilde{\Psi}^{(2hP+R_m)}_{HP}(q) - \sum_{h=0}^{ \frac{H-1}{2}-1} e^{2\pi i \frac{u}{H} (2(h+1)P-R_m)} \widetilde{\Psi}^{(2(h+1)P-R_m)}_{HP}(q) \Big) 	\label{wrt-rs-hblock2}
\end{align}
Splitting the sum into $u=0$ part and the other part $u = 1, \ldots, H-1$, \eqref{wrt-rs-hblock2} can be written as
\begin{align}
\begin{split}
&\hspace{-3.5mm}\sum_{v=0}^{s-1} e^{2\pi i \frac{r}{s} HP v^2}  \sum_{m=0}^{3} \Big(  \sum_{h=0}^{ \frac{H-1}{2}}  \widetilde{\Psi}^{(2hP+R_m)}_{HP}(q) - \sum_{h=0}^{ \frac{H-1}{2}-1}  \widetilde{\Psi}^{(2(h+1)P-R_m)}_{HP}(q) \Big) 	\\
&\hspace{-15mm}+ \sum_{u=1}^{\frac{H-1}{2}} \sum_{v=0}^{s-1} \Bigg[ 
e^{2\pi i \frac{r}{s} PH v^2}  e^{2\pi i \frac{r}{s} \frac{P}{H} u^2}  \sum_{m=0}^{3} \bigg(  
\sum_{h=0}^{ \frac{H-1}{2} } \big( e^{-4\pi i \frac{r}{s} P uv} e^{-2\pi i \frac{u}{H} (2hP+R_m)} + e^{4\pi i \frac{r}{s} P uv} e^{2\pi i \frac{u}{H} (2hP+R_m)} \big) \widetilde{\Psi}^{(2hP+R_m)}_{HP}(q)	\\
&\hspace{5mm}- \sum_{h=0}^{ \frac{H-1}{2}-1} \big( e^{-4\pi i \frac{r}{s} P uv} e^{-2\pi i \frac{u}{H} (2(h+1)P-R_m)} + e^{4\pi i \frac{r}{s} P uv} e^{2\pi i \frac{u}{H} (2(h+1)P-R_m)} \big) \widetilde{\Psi}^{(2(h+1)P-R_m)}_{HP}(q) \bigg) 	
\Bigg] \Bigg|_{q \searrow e^{2\pi i \frac{1}{K}}}	\,	.
\end{split}	\label{wrt-rs-u}
\end{align}
We can do similarly for the sum over $s$ as we did for $u$ where $s$ is chosen as an odd number here.
For the second and the third line of \eqref{wrt-rs-u}, after splitting the sum into $v=0$ part and $v=1, \ldots,  s-1$ part, we rewrite the sum over $v=1, \ldots,  s-1$ as a sum over $v=1, \ldots, \frac{s-1}{2}$.
The summand of the sum over $v=1, \ldots, \frac{s-1}{2}$ has a common factor $\big( e^{-4\pi i \frac{r}{s} P uv} + e^{4\pi i \frac{r}{s} P uv}\big)$, so the sum can be expressed as
\begin{align}
\begin{split}
&\hspace{-3mm} \sum_{u=1}^{\frac{H-1}{2}} \sum_{v=1}^{\frac{s-1}{2}} \Bigg[ 
e^{2\pi i \frac{r}{s} PH v^2}  e^{2\pi i \frac{r}{s} \frac{P}{H} u^2}  \big( e^{-4\pi i \frac{r}{s} P uv} + e^{4\pi i \frac{r}{s} P uv}\big)	\\
&\hspace{15mm}\times 
\sum_{m=0}^{3} \bigg(  
\sum_{h=0}^{ \frac{H-1}{2} } \big(  e^{-2\pi i \frac{u}{H} (2hP+R_m)} + e^{2\pi i \frac{u}{H} (2hP+R_m)} \big) \widetilde{\Psi}^{(2hP+R_m)}_{HP}(q)	\\
&\hspace{30mm}- \sum_{h=0}^{ \frac{H-1}{2}-1 } \big( e^{-2\pi i \frac{u}{H} (2(h+1)P-R_m)} + e^{2\pi i \frac{u}{H} (2(h+1)P-R_m)} \big) \widetilde{\Psi}^{(2(h+1)P-R_m)}_{HP}(q) \bigg) 	
\Bigg] \Bigg|_{q \searrow e^{2\pi i \frac{1}{K}}}	\,	.	\label{wrt-rs1-odd}
\end{split}
\end{align}
Since $\sum_{v=1}^{\frac{s-1}{2}} e^{2\pi i \frac{r}{s} PH v^2} \big( e^{-4\pi i \frac{r}{s} P uv} + e^{4\pi i \frac{r}{s} P uv}\big) = \sum_{v=1}^{s-1} e^{2\pi i \frac{r}{s} PH v^2} e^{4\pi i \frac{r}{s} P uv}$, from \eqref{wrt-rs-u} and \eqref{wrt-rs1-odd}, we obtain 
\begin{align}
\begin{split}
&\hspace{-3mm}Z_{K=\frac{r}{s}}(M_3) = \sum_{v=0}^{s-1} e^{2\pi i \frac{r}{s} HP v^2}  \sum_{m=0}^{3} \Big(  \sum_{h=0}^{ \frac{H-1}{2} }  \widetilde{\Psi}^{(2hP+R_m)}_{HP}(q) - \sum_{h=0}^{ \frac{H-1}{2}-1 }  \widetilde{\Psi}^{(2(h+1)P-R_m)}_{HP}(q) \Big)	\\
&\hspace{4mm}+ \sum_{u=1}^{\frac{H-1}{2}} \sum_{v=0}^{s-1} \Bigg[ 
e^{2\pi i \frac{r}{s} \frac{P}{H} (vH+u)^2} 
\sum_{m=0}^{3} \bigg(  
\sum_{h=0}^{ \frac{H-1}{2} } \big(  e^{-2\pi i \frac{u}{H} (2hP+R_m)} + e^{2\pi i \frac{u}{H} (2hP+R_m)} \big) \widetilde{\Psi}^{(2hP+R_m)}_{HP}(q)	\\
&\hspace{35mm}- \sum_{h=0}^{ \frac{H-1}{2}-1 } \big( e^{-2\pi i \frac{u}{H} (2(h+1)P-R_m)} + e^{2\pi i \frac{u}{H} (2(h+1)P-R_m)} \big) \widetilde{\Psi}^{(2(h+1)P-R_m)}_{HP}(q) \bigg) 	
\Bigg]	\Bigg|_{q \searrow e^{2\pi i \frac{1}{K}}}	\,	.	\label{wrt-rs-fin-odd}
\end{split}	
\end{align}

When $H$ is even, we also have similar results,
\begin{align}
\begin{split}
&\hspace{-3mm} Z_{K=\frac{r}{s}}(M_3) 
=  \sum_{v=0}^{s-1} e^{2\pi i \frac{r}{s} HP v^2}  \sum_{m=0}^{3} \sum_{h=0}^{ \frac{H}{2}-1 } \Big(   \widetilde{\Psi}^{(2hP+R_m)}_{HP}(q) -  \widetilde{\Psi}^{(2(h+1)P-R_m)}_{HP}(q) \Big)	\\
&\hspace{3mm}+ \sum_{u=1}^{\frac{H}{2}-1} \sum_{v=0}^{s-1} \Bigg[ 
e^{2\pi i \frac{r}{s} \frac{P}{H} (vH+u)^2} 
\sum_{m=0}^{3} \sum_{h=0}^{ \frac{H}{2}-1 } \bigg(  
 \big(  e^{-2\pi i \frac{u}{H} (2hP+R_m)} + e^{2\pi i \frac{u}{H} (2hP+R_m)} \big) \widetilde{\Psi}^{(2hP+R_m)}_{HP}(q)	\\
&\hspace{55mm} -  \big( e^{-2\pi i \frac{u}{H} (2(h+1)P-R_m)} + e^{2\pi i \frac{u}{H} (2(h+1)P-R_m)} \big) \widetilde{\Psi}^{(2(h+1)P-R_m)}_{HP}(q) \bigg) 	
\Bigg]		\\
&\hspace{3mm}+ \sum_{v=0}^{s-1} 
e^{2\pi i \frac{r}{s} PH (v+\frac{1}{2})^2} 
\sum_{m=0}^{3} \sum_{h=0}^{ \frac{H}{2}-1 } \bigg(  
 e^{ \pi i (2hP+R_m)} \widetilde{\Psi}^{(2hP+R_m)}_{HP}(q) - e^{ \pi i (2(h+1)P-R_m)} \widetilde{\Psi}^{(2(h+1)P-R_m)}_{HP}(q) \bigg) 	\Bigg|_{q \searrow e^{2\pi i \frac{1}{K}}}	\label{wrt-rs-fin-even}
\end{split}	
\end{align}
where the last line is from $u=\frac{H}{2}$.

Comparing with the case of the standard root of unity, the factor $\sum_{v=0}^{s-1} e^{2\pi i \frac{r}{s} \frac{P}{H} (Hv+u)^2}$ is different. 
Other than that, the structure is the same as in the case of the standard root of unity.
As a consistency check, when $s=1$ we obtain the result for the standard root of unity.

%%%%%%%%%%%%%%%%%%%%%%%%%%%%%%%%%%%%%%%%%%%%%%%%%%%%%%%%%%%%%%%%%%%%%%%%%%%%%%%%%%%%%%

\subsubsection*{Remarks on the larger number of singular fibers}

As discussed in \cite{Chung-wrt} for an integer $K$, the calculation for the case of larger number of singular fibers $F \geq 4$ is parallel to the case of $F=3$, so the structures are the same.
Therefore, from the calculation above, when $F\geq 4$ and $K$ is a rational number, the structure of the expression is also the same as in the case of $F=3$ above.
Thus, for the case of a rational $K$ and $F \geq 4$, we can just use the formulas in Appendix A and B in \cite{Chung-wrt}, replace $e^{2\pi i K \frac{P}{H}t^2}$ with $\sum_{v=0}^{s-1} e^{2\pi i \frac{r}{s} \frac{P}{H} (Hv+u)^2}$, and take a limit $q \searrow e^{2\pi i \frac{s}{r}}$.\footnote{When $F\geq 4$, the homological block is given by $\widetilde{\Phi}(q)$, a false theta function that is the Eichler integral of the modular form $\Phi^{(l)}_P(q) := \sum_{n=0}^{\infty} \phi_{2P}^{(l)}(n) q^{\frac{n^2}{4P}}$ of half-integer weight $1/2$, and their derivatives.
Here, $\phi_{2P}^{(l)}(n)$ is a periodic function, which is $1$ when $n=\pm l$ mod $2P$ and zero otherwise.
The limit $q \searrow e^{2\pi i\frac{s}{r}}$ of them and their derivatives have been discussed in \cite{Hikami-lattice1, Hikami-lattice2}. }

%%%%%%%%%%%%%%%%%%%%%%%%%%%%%%%%%%%%%%%%%%%%%%%%%%%%%%%%%%%%%%%%%%%%%%%%%%%%%%%%%%%%%%
%%%%%%%%%%%%%%%%%%%%%%%%%%%%%%%%%%%%%%%%%%%%%%%%%%%%%%%%%%%%%%%%%%%%%%%%%%%%%%%%%%%%%%
%%%%%%%%%%%%%%%%%%%%%%%%%%%%%%%%%%%%%%%%%%%%%%%%%%%%%%%%%%%%%%%%%%%%%%%%%%%%%%%%%%%%%%

\subsection{Properties and general structure}

In the case of the standard root of unity, the summation variable $t=0,1, \ldots, H-1$ up to the Weyl group action $t \leftrightarrow -t$ mod $H$ corresponds to an abelian flat connection whose holonomy for the central element of $\pi_1(M_3)$ is a conjugacy class of $\text{diag} \, (e^{2 \pi i \frac{P}{H}t}, e^{-2 \pi i \frac{P}{H}t})$.
In particular, $t=0$ corresponds to the trivial flat connection.

In the case of other roots of unity $K=\frac{r}{s}$ where $s$ is not equal to 1, $t$ goes from $0$ to $Hs-1$.
But as discussed in section \ref{ssec:h2}, it was possible to express \eqref{wrt-rs-hblock} as \eqref{wrt-rs-fin-odd} or \eqref{wrt-rs-fin-even} that are expressed as a sum over $u=0,1, \ldots, H-1$ and $v=0,1, \ldots, s-1$.
And we saw that, other than the limit $q\searrow e^{2\pi i \frac{s}{r}}$, the difference from the case of the standard root of unity is $\sum_{v=0}^{s-1} e^{2\pi i \frac{r}{s} \frac{P}{H} (Hv+u)^2}$ while it was just $e^{2\pi i r \frac{P}{H} u^2}$ when $s=1$.
That is, given that $Z_{K}(M_3) = \sum_{t} e^{2\pi i K\frac{P}{H} t^2} Z_t \big|_{q\searrow e^{\frac{2\pi i}{K}}}$ schematically in the case of the standard root of unity where $Z_t$ is the contribution from the abelian flat connection $t$, when $K=\frac{r}{s}$ with $r$ and $s$ being coprime integers and $s\neq1$, \eqref{wrt-rs-fin-odd} and \eqref{wrt-rs-fin-even} indicate that $Z_{K}(M_3) = \sum_{v=0}^{s-1} \sum_{u} e^{2\pi i \frac{r}{s} \frac{P}{H} (Hv+u)^2} Z_u \big|_{q\searrow e^{2\pi i \frac{s}{r}}}$.

Upon $u \rightarrow u+H$, $\sum_{v=0}^{s-1} e^{2\pi i \frac{r}{s} \frac{P}{H} (Hv+u)^2}$ becomes $\sum_{v=0}^{s-1} e^{2\pi i \frac{r}{s} \frac{P}{H} (H(v+1)+u)^2}$, but one can see that it is the same with $\sum_{v=0}^{s-1} e^{2\pi i \frac{r}{s} \frac{P}{H} (Hv+u)^2}$, and also the other summand stays the same.
Therefore, the expression is invariant under $u \rightarrow u+H$.
In addition, upon $u \rightarrow -u$, the summand in \eqref{wrt-rs-fin-odd} and \eqref{wrt-rs-fin-even} are the same, but $\sum_{v=0}^{s-1} e^{2\pi i \frac{r}{s} \frac{P}{H} (vH+u)^2}$ becomes $\sum_{v=0}^{s-1} e^{2\pi i \frac{r}{s} \frac{P}{H} (vH-u)^2}$.
However, from $\sum_{v=0}^{s-1} e^{2\pi i \frac{r}{s} \frac{P}{H} (Hv-u)^2} = e^{2\pi i \frac{r}{s} \frac{P}{H} u^2} + \sum_{v=1}^{s-1} e^{2\pi i \frac{r}{s} \frac{P}{H} (Hv-u)^2}$, by  rewriting the RHS in terms of $v'=s-v$ and renaming $v'$ to $v$, the RHS becomes to $\sum_{v=0}^{s-1} e^{2\pi i \frac{r}{s} \frac{P}{H} (Hv+u)^2}$.
Thus, the expression is invariant under $u \leftrightarrow -u$ and this is consistent with that
the Weyl orbits of $(u,-u)$ mod $H$ correspond to the same abelian flat connection.

Also, $u \leftrightarrow -u$ can be regarded as a complex conjugate of $(u,-u)$ at the level of holonomy.
Therefore, as in the case of the standard root of unity, we can say that the contributions from the abelian flat connnection and from the conjugate abelian flat connection are the same, though in the case of $SU(2)$ these two abelian flat connections are equivalent from the beginning so that contributions from them are obviously the same.	\\

In addition, as in the case of the standard root of unity, the contributions from the abelian flat connections that are related by the action of the center of $SU(2)$ are the same up to an overall factor $e^{\pi i r}$ and this happens when $H$ is a multiple of 2 but not of 4 \cite{Chung-wrt, Cheng:2018vpl}.
More precisely, from the $\sum_{v=0}^{s-1} e^{2\pi i K \frac{P}{H} (vH+u)^2}$ factor, we see from a number of examples where $s$ is odd that when $H$ is a multiple of 2 but not of 4, an additional factor $e^{\pi i r}$ arises for the abelian flat connection $(u+\frac{H}{2}, -u-\frac{H}{2})$ compared to the case of $(u,-u)$.
There is no such factor when $H$ is a multiple of 4.
Hence, the contributions from abelian flat connections that are related by the action of the center can have a different factor by $e^{\pi i r}$ depending on $H$.	\\

%%%%%%%%%%%%%%%%%%%%%%%%%%%%%%%%%%%%%%%%%%%%%%%%%%%%%%%%%%%%%%%%%%%%%%%%%%%%%%%%%%%%%%

\subsubsection*{General structure}

We use similar notations as in \cite{Chung-wrt}.
The Weyl orbit of $(u,-u) \in (\mathbb{Z}_H)^2/\mathbb{Z}_2$ is denoted as $W_{u}$.
When $H$ is even, since Weyl orbits $W_u$ and $W_{u+\frac{H}{2}}$ that are related by the action of the center give the same contribution to the WRT invariant up to an overall factor $e^{\pi i r}$, we group $W_u$ and $W_{u+\frac{H}{2}}$ by orbits under the action of the center, which we denote by $C_a$ where $a$ is a label for an abelian flat connection.
The range of $a$ is $a=0,1, \ldots, \frac{H-2}{4}$ when $H$ is a multiple of 2 but not of 4, and $a=0,1, \ldots, \frac{H}{4}$ when $H$ is a multiple of 4.
We denote elements in the Weyl orbit $W_u$ by $\tilde{u}$ and a representative of any elements of $W_u$ in $C_b$ by $\tilde{b}$.

With the notation above, the $S$-matrix is the same as the one in the case of the standard root of unity
\begin{align}
S_{ab}= \frac{1}{ \sqrt{\text{gcd}(2,H)} }\sum_{W_u \in C_a} \frac{ \sum_{\tilde{u} \in W_u} e^{2\pi i  lk (\tilde{u},\tilde{b})}}{ |\text{Tor} \, H_1(M_3,\mathbb{Z})|^{\frac{1}{2}}}	\label{su2-smat0}
\end{align}
with
\begin{align}
lk(u,u') = \frac{P}{H} \sum_{j=1}^2 u_j u'_j = \frac{2P}{H} u_1 u'_1	\label{su2-lk}
\end{align}
where $u=(u_1,-u_1)$ and $u'=(u'_1,-u'_1)$.
We often use the notation $lk(a,b) := lk(\tilde{a}, \tilde{b})$.

Then, the WRT invariant is given by
\begin{align}
Z_{K=\frac{r}{s}}(M_3) = \frac{B}{2i} q^{-\phi/4} \bigg( \frac{2i}{r/s} \frac{P}{H} \bigg)^{1/2} \sqrt{\text{gcd} (2,H) H} \sum_{v=0}^{s-1} \sum_{a,b} e^{2\pi i \frac{r}{s} \frac{P}{H} (Hv+a)^2 } S_{ab} \widehat{Z}_b(q)	\,	\Big|_{q \searrow e^{2\pi i \frac{s}{r}}}		\label{su2-wrt0}
\end{align}
when $H$ is odd or a multiple of 4.

When $H$ is a multiple of 2 but not of 4, the WRT invariant can be written as
\begin{align}
Z_{K=\frac{r}{s}}(M_3) = \frac{B}{2i} q^{-\phi/4} \bigg( \frac{2i}{r/s} \frac{P}{H} \bigg)^{1/2} \sqrt{ \frac{H}{2}} \sum_{v=0}^{s-1} \sum_{\dot{a},\dot{b}} e^{2 \pi i \frac{r}{s} \frac{P}{H} (Hv+\dot{a})^2} (Y \otimes S_{ab})_{\dot{a}\dot{b}} \widehat{Z}_{\dot{b}}(q)	\,	\Big|_{q \searrow e^{2\pi i \frac{s}{r}}}	
\label{su2-wrt1}
\end{align}
where $\dot{a}, \dot{b}= \dot{0}, \ldots, \dot{\frac{H}{2}}$, $Y= \begin{pmatrix} 1 & 1 \\ 1 & 1 \end{pmatrix}$ so that $(Y \otimes S_{ab})_{\dot{a}\dot{b}}=\begin{pmatrix} S_{ab} & S_{ab} \\ S_{ab} & S_{ab} \end{pmatrix}$, and $e^{2 \pi i \frac{r}{s} \frac{P}{H} (Hv+\dot{a})^2} = (e^{2 \pi i \frac{r}{s} \frac{P}{H} (Hv+\dot{0})^2}, \ldots, e^{2 \pi i \frac{r}{s} \frac{P}{H} (Hv+ \dot{\frac{H-2}{4}})^2}, e^{2 \pi i \frac{r}{s} \frac{P}{H} (Hv+\dot{0})^2} e^{\pi i r}, \ldots, e^{2 \pi i \frac{r}{s} \frac{P}{H} (Hv+ \dot{\frac{H-2}{4}})^2} e^{\pi i r})$.
Since $Z_a = Z_{\frac{H}{2}-a}$ for $a=0,1, \ldots, \frac{H-2}{4}$, in the notation of \eqref{su2-wrt1} we have $Z_{\dot{a}}(q) = Z_{\dot{a + \frac{H+2}{4}}}(q)$, $\dot{a}=\dot{0}, \ldots, \dot{\frac{H-2}{4}}$ with $Z_{\dot{a}}(q) = (Y \otimes S_{ab})_{\dot{a}\dot{b}} \widehat{Z}_{\dot{b}}(q)$ where $Z_{\dot{a}}(q) = Z_{a}(q)$ for $a=0,1, \ldots, \frac{H-2}{4}$ and $Z_{\dot{a}}(q) = Z_{\frac{3H+2}{4}-a}(q)$ for $a= \frac{H}{2}, \frac{H}{2}-1, \ldots, \frac{H+2}{4}$.
Similarly, we denote $\widehat{Z}_{\dot{b}}(q) = \widehat{Z}_{b}(q)$ for $b=0,1, \ldots, \frac{H-2}{4}$ and $\widehat{Z}_{\dot{b}}(q) = \widehat{Z}_{\frac{3H+2}{4}-b}(q)$ for $b= \frac{H}{2}, \frac{H}{2}-1, \ldots, \frac{H+2}{4}$.

We expect that \eqref{su2-wrt0} and \eqref{su2-wrt1} hold for general rational homology spheres.

%%%%%%%%%%%%%%%%%%%%%%%%%%%%%%%%%%%%%%%%%%%%%%%%%%%%%%%%%%%%%%%%%%%%%%%%%%%%%%%%%%%%%%
%%%%%%%%%%%%%%%%%%%%%%%%%%%%%%%%%%%%%%%%%%%%%%%%%%%%%%%%%%%%%%%%%%%%%%%%%%%%%%%%%%%%%%
%%%%%%%%%%%%%%%%%%%%%%%%%%%%%%%%%%%%%%%%%%%%%%%%%%%%%%%%%%%%%%%%%%%%%%%%%%%%%%%%%%%%%%

\subsection{Examples}
In this section, we provide some examples.
We mostly quote examples from \cite{Chung-wrt}.

%%%%%%%%%%%%%%%%%%%%%%%%%%%%%%%%%%%%%%%%%%%%%%%%%%%%%%%%%%%%%%%%%%%%%%%%%%%%%%%%%%%%%%

\subsubsection*{\textbullet \ $(P_1,P_2,P_3)=(3,5,7)$ with $H=2$}

For the case of the standard root of unity $q=e^{\frac{2\pi i}{K}}$, $K \in \mathbb{Z}$, the WRT invariant can be expressed as 
\begin{align}
Z_{K}(M_3) = \frac{B}{2i} q^{-\phi_3/4} \bigg( \frac{105i}{K}  \bigg)^{1/2} (1+ e^{\pi i K}) (\widehat{Z}_0 + \widehat{Z}_1)	\,	\bigg|_{q \searrow e^{2\pi i \frac{1}{K}}}	
\end{align}
where 
\begin{align}
\widehat{Z}_0(q) &= \widetilde{\Psi }_{210}^{(34)}+\widetilde{\Psi }_{210}^{(106)}+\widetilde{\Psi }_{210}^{(134)}+\widetilde{\Psi }_{210}^{(146)}	\,	,	\\
\widehat{Z}_1(q) &= -\widetilde{\Psi }_{210}^{(64)}-\widetilde{\Psi }_{210}^{(76)}-\widetilde{\Psi }_{210}^{(104)}-\widetilde{\Psi }_{210}^{(176)}	\,	.
\end{align}
For other roots of unity, we have 
\begin{align}
Z_{K=\frac{r}{s}}(M_3) = \frac{B}{2i} q^{-\phi_3/4} \bigg( \frac{105i}{r/s}  \bigg)^{1/2} \omega \, (1+ e^{\pi i r}) (\widehat{Z}_0 + \widehat{Z}_1)	\,	\bigg|_{q \searrow e^{2\pi i \frac{s}{r}}}	\,	.	\label{exh2}
\end{align}
For example, when $K=\frac{r}{s} = \frac{2}{11}, \frac{4}{11}, \frac{7}{13}, \frac{5}{17}$, $\omega = -i, i,1,1 $, respectively, where \eqref{exh2} vanishes when $r$ is odd.
Or we may put it in the form of
\begin{align}
Z_{K=\frac{r}{s}}(M_3) = \frac{B}{2i} q^{-\phi_3/4} \bigg( \frac{105i}{r/s}  \bigg)^{1/2} \sum_{a,b=0}^{1} \omega_a Y_{ab} \widehat{Z}_b	\,	\bigg|_{q \searrow e^{2\pi i \frac{s}{r}}}		\,	,
\end{align}
where $Y_{ab} = \begin{pmatrix} 1 & 1 \\ 1 & 1 \end{pmatrix}$ and $(w_0,w_1) = (\omega,\omega e^{\pi i r})$.	\\

%%%%%%%%%%%%%%%%%%%%%%%%%%%%%%%%%%%%%%%%%%%%%%%%%%%%%%%%%%%%%%%%%%%%%%%%%%%%%%%%%%%%%%

\subsubsection*{\textbullet \ $(P_1,P_2,P_3)=(2,5,7)$ with $H=3$}

The WRT invariant at the standard root of unity can be written as
\begin{align}
\begin{split}
Z_{K}(M_3)
=& \frac{B}{i} q^{-\phi_3/4} \bigg( \frac{70i}{K} \bigg)^{1/2} 
\sum_{a,b=0}^{1} e^{2 \pi i K CS_a }S_{ab} \widehat{Z}_{b}	\,	\bigg|_{q \searrow e^{2\pi i \frac{1}{K}}}		\,	,
\end{split}
\end{align}
where
\begin{align}
\widehat{Z}_0 &= 
-\widetilde{\Psi }_{210}^{(39)}-\widetilde{\Psi }_{210}^{(81)}-\widetilde{\Psi }_{210}^{(129)}+\widetilde{\Psi }_{210}^{(249)}	\,	,	\\
\begin{split}
\widehat{Z}_1 &= 
\widetilde{\Psi }_{210}^{(11)}-\widetilde{\Psi }_{210}^{(31)}+\widetilde{\Psi }_{210}^{(59)}+\widetilde{\Psi }_{210}^{(101)}+\widetilde{\Psi }_{210}^{(109)}+\widetilde{\Psi}_{210}^{(151)}+\widetilde{\Psi }_{210}^{(199)}+\widetilde{\Psi }_{210}^{(241)}	\,	,
\end{split}
\end{align}
$(CS_0,CS_1)=(0,\frac{1}{3})$, and the $S$-matrix is
\begin{align}
S_{ab} =\frac{1}{\sqrt{3}} 
\begin{pmatrix}
1	&	1	\\
2	&	-1
\end{pmatrix}	\,	.	\label{s-mat-su2-h3}
\end{align}
For other roots of unity, we have
\begin{align}
\begin{split}
Z_{K=\frac{r}{s}}(M_3) 
=& \frac{B}{i} q^{-\phi_3/4} \bigg( \frac{70i}{r/s} \bigg)^{1/2} 
\sum_{a,b=0}^{1} \omega_a S_{ab} \widehat{Z}_{b}	\,	\bigg|_{q \searrow e^{2\pi i \frac{s}{r}}}		\,	.
\end{split}
\end{align}
For example, when $K=\frac{r}{s} = \frac{5}{11}, \frac{7}{11}, \frac{11}{13}, \frac{15}{13}$, $(\omega_1,\omega_2) = (i, e^{\frac{7 \pi i}{6}}), (-i,e^{\frac{5\pi i}{6}}), (1,e^{\frac{4\pi i}{3}}), (1,1)$, respectively.

%%%%%%%%%%%%%%%%%%%%%%%%%%%%%%%%%%%%%%%%%%%%%%%%%%%%%%%%%%%%%%%%%%%%%%%%%%%%%%%%%%%%%%

\subsubsection*{\textbullet \ $(P_1,P_2,P_3)=(3,5,7)$ with $H=4$}

The WRT invariant with $K \in \mathbb{Z}$ is given by
\begin{align}
\begin{split}
Z_{K}(M_3) =
\frac{B}{2i} q^{-\phi_3/4} \bigg( \frac{105i}{2K} \bigg)^{1/2} 
\big( Z_0 + e^{\frac{\pi i}{2} K} Z_1 + Z_2 \big)	=
\frac{B}{i} q^{-\phi_3/4} \bigg( \frac{105 i}{K} \bigg)^{1/2} \sum_{a,b=0}^{1} e^{2 \pi i K CS_a} S_{ab} \widehat{Z}_{b}	\,	\bigg|_{q \searrow e^{2\pi i \frac{1}{K}}}		\,	,
\label{3f-h4-result0}
\end{split}
\end{align}
where $Z_0 = Z_2$, $(CS_0,CS_1)=(0,\frac{1}{4})$, and
\begin{align}
\begin{split}
\widehat{Z}_0 =
-\widetilde{\Psi }_{420}^{(64)}-\widetilde{\Psi }_{420}^{(76)}-\widetilde{\Psi }_{420}^{(104)}-\widetilde{\Psi }_{420}^{(176)}+\widetilde{\Psi }_{420}^{(244)}+\widetilde{\Psi
   }_{420}^{(316)}+\widetilde{\Psi }_{420}^{(344)}+\widetilde{\Psi }_{420}^{(356)}	\,	,
\end{split}	\\
\begin{split}
\widehat{Z}_1 =
\widetilde{\Psi }_{420}^{(34)}+\widetilde{\Psi }_{420}^{(106)}+\widetilde{\Psi }_{420}^{(134)}+\widetilde{\Psi }_{420}^{(146)}-\widetilde{\Psi }_{420}^{(274)}-\widetilde{\Psi
   }_{420}^{(286)}-\widetilde{\Psi }_{420}^{(314)}-\widetilde{\Psi }_{420}^{(386)}	\,	.
\end{split}
\end{align}
$Z_a$ and $\widehat{Z}_b$ are related by $Z_0 = \widehat{Z}_0 + \widehat{Z}_1$ and $\frac{1}{2}Z_1 = \widehat{Z}_0 - \widehat{Z}_1$, so 
the $S$-matrix is given by
\begin{align}S_{ab} = 
\frac{1}{\sqrt{2}}\begin{pmatrix}
1	&	1	\\	1	&	-1
\end{pmatrix}	\,	.	\label{s-mat-su2-h4}
\end{align}
The WRT invariant at other roots of unity is given by
\begin{align}
\begin{split}
Z_{K=\frac{r}{s}}(M) 
=\frac{B}{i} q^{-\phi_3/4} \bigg( \frac{105 i}{r/s} \bigg)^{1/2} \sum_{a,b=0}^{1} \omega_a S_{ab} \widehat{Z}_{b}	\,	\bigg|_{q \searrow e^{2\pi i \frac{s}{r}}}		\,	.
\label{3f-h4-result1}
\end{split}
\end{align}
For example, when $K=\frac{2}{11}, \frac{3}{11}, \frac{7}{13}, \frac{4}{17}$, $(\omega_0,\omega_1) = (i,-i), (-i,1), (1,-1), (-1,-1)$, respectively.

%%%%%%%%%%%%%%%%%%%%%%%%%%%%%%%%%%%%%%%%%%%%%%%%%%%%%%%%%%%%%%%%%%%%%%%%%%%%%%%%%%%%%%

\subsubsection*{\textbullet \ $(P_1,P_2,P_3)=(2,3,11)$ with $H=5$}

The WRT invariant with an integer $K \in \mathbb{Z}$ is given by 
\begin{align}
Z_{K}(M_3) =  \frac{B}{i} q^{-\phi/4} \bigg( \frac{33i}{K}  \bigg)^{1/2} \sum_{a,b=0}^{2} e^{2 \pi i K CS_a} S_{ab} \widehat{Z}_b	\bigg|_{q \searrow e^{2\pi i \frac{1}{K}}}
\end{align}
where $K$ is an integer and $(CS_0,CS_1,CS_2)=(0,\frac{1}{5}, \frac{4}{5})$.
The homological blocks are
\begin{align}
\widehat{Z}_0 &= \tilde{\Psi }_{330}^{(5)}+\tilde{\Psi }_{330}^{(115)}+\tilde{\Psi}_{330}^{(215)}+\tilde{\Psi }_{330}^{(325)}	\\
\widehat{Z}_1 &= -\tilde{\Psi }_{330}^{(17)}+\tilde{\Psi }_{330}^{(83)}-\tilde{\Psi}_{330}^{(127)}+\tilde{\Psi }_{330}^{(137)}+\tilde{\Psi}_{330}^{(193)}-\tilde{\Psi }_{330}^{(203)}+\tilde{\Psi}_{330}^{(247)}+\tilde{\Psi }_{330}^{(347)}	\\
\widehat{Z}_2 &= -\tilde{\Psi }_{330}^{(49)}+\tilde{\Psi }_{330}^{(61)}-\tilde{\Psi}_{330}^{(71)}-\tilde{\Psi }_{330}^{(149)}-\tilde{\Psi}_{330}^{(181)}-\tilde{\Psi }_{330}^{(259)}+\tilde{\Psi}_{330}^{(269)}+\tilde{\Psi }_{330}^{(379)}
\end{align}
and the $S$-matrix is
\begin{align}
S_{ab} = \frac{1}{\sqrt{5}} \begin{pmatrix} 1 & 1 & 1 \\ 2 & A & B \\ 2 & B & A \end{pmatrix}	\,	
\end{align}
where $A=\frac{1}{2} \left(-\sqrt{5}-1\right)$ and $\frac{1}{2} \left(\sqrt{5}-1\right)$.
Thus, we have
\begin{align}
Z_{K=\frac{r}{s}}(M_3)  =\frac{B}{i} q^{-\phi/4} \bigg( \frac{33i}{r/s}  \bigg)^{1/2} 
\sum_{a,b=0}^{2} \omega_a S_{ab} \widehat{Z}_b	\bigg|_{q \searrow e^{2\pi i \frac{s}{r}}}	\,	.
\end{align}
For example, when $K=\frac{r}{s} =\frac{2}{13}, \frac{13}{17}, \frac{17}{19}, \frac{11}{5}$, $(\omega_1, \omega_2, \omega_3)$ $= (1, e^{\frac{8\pi i}{5}}, e^{\frac{2\pi i}{5}})$, $(-1,e^{\frac{3\pi i}{5}}, e^{\frac{7\pi i}{5}})$, $(i, e^{\frac{17\pi i}{10}}, e^{\frac{13\pi i}{10}})$, $(\sqrt{5},0,0)$, respectively.

%%%%%%%%%%%%%%%%%%%%%%%%%%%%%%%%%%%%%%%%%%%%%%%%%%%%%%%%%%%%%%%%%%%%%%%%%%%%%%%%%%%%%%

\subsubsection*{\textbullet \ $(P_1,P_2,P_3)=(5,7,11)$ with $H=6$}

The WRT invariant with $K \in \mathbb{Z}$ is given by 
\begin{align}
Z_{K}(M_3) 
=& \frac{B}{2i} q^{-\phi_3/4} \bigg( \frac{385i}{3K} \bigg)^{1/2} 
\Big( Z_0 + e^{2 \pi i K \frac{1}{6}} Z_1 + e^{ \pi i K} Z_2 + e^{2 \pi i K \frac{4}{6}} Z_3 \Big)	\\
=& \frac{B}{2i} q^{-\phi_3/4} \bigg(\frac{385i}{K} \bigg)^{1/2} 
\sum_{\dot{a}, \dot{b}=\dot{0}}^{\dot{3}} e^{2 \pi i K CS_{\dot{a}}} \Big( \begin{pmatrix} 1 & 1 \\ 1 & 1 \end{pmatrix} \otimes S_{ab} \Big)_{\dot{a} \dot{b}} \widehat{Z}_{\dot{b}}(q) 	\label{3f-h6-result00}
\end{align}
where $Z_0=Z_3$, $Z_1=Z_2$, $\widehat{Z}_{\dot{b}}=(\widehat{Z}_{0},\widehat{Z}_{1},\widehat{Z}_{3},\widehat{Z}_{2})^{T}$ and $CS_{\dot{a}} = (CS_0, CS_1, CS_3, CS_2) = (0, \frac{1}{6}, \frac{1}{2}, \frac{2}{3})$.
Homological blocks are given by
\begin{align}
\widehat{Z}_0 &= -\widetilde{\Psi }_{2310}^{(288)}-\widetilde{\Psi }_{2310}^{(372)}-\widetilde{\Psi }_{2310}^{(552)}+\widetilde{\Psi }_{2310}^{(1212)}	\,	,	\\
\widehat{Z}_1 &= -\widetilde{\Psi }_{2310}^{(328)}+\widetilde{\Psi }_{2310}^{(988)}+\widetilde{\Psi }_{2310}^{(1168)}+\widetilde{\Psi }_{2310}^{(1252)}-\widetilde{\Psi}_{2310}^{(1828)}-\widetilde{\Psi }_{2310}^{(1868)}-\widetilde{\Psi }_{2310}^{(1912)}-\widetilde{\Psi }_{2310}^{(2092)}	\,	,	\\
\widehat{Z}_2 &= \widetilde{\Psi }_{2310}^{(218)}+\widetilde{\Psi }_{2310}^{(398)}+\widetilde{\Psi }_{2310}^{(442)}+\widetilde{\Psi }_{2310}^{(482)}-\widetilde{\Psi}_{2310}^{(1058)}-\widetilde{\Psi }_{2310}^{(1142)}-\widetilde{\Psi }_{2310}^{(1322)}+\widetilde{\Psi }_{2310}^{(1982)}	\,	\\
\widehat{Z}_3 &= -\widetilde{\Psi }_{2310}^{(1098)}+\widetilde{\Psi }_{2310}^{(1758)}+\widetilde{\Psi }_{2310}^{(1938)}+\widetilde{\Psi }_{2310}^{(2022)}	\,	,
\end{align}
which are related to $Z_a$'s as $Z_0 = Z_3 = \widehat{Z}_0 + \widehat{Z}_1 + \widehat{Z}_2 + \widehat{Z}_3$ and $Z_1 = Z_2 = 2 \widehat{Z}_0 - \widehat{Z}_1 +2 \widehat{Z}_3 - \widehat{Z}_2$.

When $K$ is rational, we have
\begin{align}
Z_{K=\frac{r}{s}}(M_3) 
=& \frac{B}{2i} q^{-\phi_3/4} \bigg( \frac{385i}{3r/s} \bigg)^{1/2} 
\sum_{a=0}^3 \omega_a Z_a	\\
=& \frac{B}{2i} q^{-\phi_3/4} \bigg(\frac{385i}{K} \bigg)^{1/2} 
\sum_{\dot{a}, \dot{b}=\dot{0}}^{\dot{3}} \omega_{\dot{a}} \Big( \begin{pmatrix} 1 & 1 \\ 1 & 1 \end{pmatrix} \otimes S_{ab} \Big)_{\dot{a} \dot{b}} \widehat{Z}_{\dot{b}}(q) 	\label{3f-h6-result0}
\end{align}
where $\omega_{\dot{a}}=(\omega_0,\omega_1,\omega_3,\omega_2)$ where $\omega_0=e^{\pi i r} \omega_3$ and $\omega_1=e^{\pi i r} \omega_2$.
For example, when $K=\frac{r}{s}=\frac{4}{13}, \frac{7}{13}, \frac{13}{3}, \frac{23}{19}$, $(\omega_0, \omega_1) = (1,e^{\frac{4\pi i}{3}}), (-1, e^{\frac{4\pi i}{3}}), (\sqrt{3},0),(i, e^{\frac{\pi i}{6}})$, respectively, where \eqref{3f-h6-result0} vanishes when $r$ is odd.

%%%%%%%%%%%%%%%%%%%%%%%%%%%%%%%%%%%%%%%%%%%%%%%%%%%%%%%%%%%%%%%%%%%%%%%%%%%%%%%%%%%%%%
%%%%%%%%%%%%%%%%%%%%%%%%%%%%%%%%%%%%%%%%%%%%%%%%%%%%%%%%%%%%%%%%%%%%%%%%%%%%%%%%%%%%%%
%%%%%%%%%%%%%%%%%%%%%%%%%%%%%%%%%%%%%%%%%%%%%%%%%%%%%%%%%%%%%%%%%%%%%%%%%%%%%%%%%%%%%%
%%%%%%%%%%%%%%%%%%%%%%%%%%%%%%%%%%%%%%%%%%%%%%%%%%%%%%%%%%%%%%%%%%%%%%%%%%%%%%%%%%%%%%
%%%%%%%%%%%%%%%%%%%%%%%%%%%%%%%%%%%%%%%%%%%%%%%%%%%%%%%%%%%%%%%%%%%%%%%%%%%%%%%%%%%%%%

\section{Asymptotic expansion of knot invariants at roots of unity}
\label{asymp-exp}

There are some versions of the volume conjecture, such as the parametrized volume conjecture and quantum volume conjecture.\footnote{See \cite{DG-review} for the review.}
Or it can also be generalized to the case with the parameter $t$ for the grading of the categorification or with the parameter $a$ in the HOMFLY polynomial.

The original volume conjecture was considered in the limit that $K \rightarrow \infty$ where the Chern-Simons level $K$ and the color $n$ of the Jones polynomial is identified $K=n$.
In the parametrized volume conjecture, both $K$ and $n$ are taken to infinity $K,n \rightarrow \infty$ while the ratio $n/K$ is fixed.
The quantum invariants are expressed in terms of $q=e^{\frac{2\pi i}{K}}$, and $q \rightarrow 1$ in the limit $K \rightarrow \infty$.

In this section, we consider an asymptotic expansion of a few knot polynomials in the limit that $q$ goes to a general roots of unity
\begin{align}
q \, \rightarrow \, \zeta^s_r := e^{2\pi i \frac{s}{r}}
\end{align} 
with $\frac{s}{r} \in \mathbb{Q}$ where $r$ and $s$ are coprime integers.

%%%%%%%%%%%%%%%%%%%%%%%%%%%%%%%%%%%%%%%%%%%%%%%%%%%%%%%%%%%%%%%%%%%%%%%%%%%%%%%%%%%%%%
%%%%%%%%%%%%%%%%%%%%%%%%%%%%%%%%%%%%%%%%%%%%%%%%%%%%%%%%%%%%%%%%%%%%%%%%%%%%%%%%%%%%%%
%%%%%%%%%%%%%%%%%%%%%%%%%%%%%%%%%%%%%%%%%%%%%%%%%%%%%%%%%%%%%%%%%%%%%%%%%%%%%%%%%%%%%%

\subsubsection*{Limit of $q$ to roots of unity}

The limit for the asymptotic expansion of the invariants that we discuss in this section is different from the standard $q\rightarrow 1$ limit with the level $K \rightarrow \infty$.
In the standard volume conjecture, the expansion parameter is $\epsilon = \frac{2\pi i}{K}$ and $\epsilon \rightarrow 0$ or $K \rightarrow \infty$ is taken.
However, when we consider the asymptotic expansion around general roots of unity $e^{2\pi i \frac{1}{K}} = e^{2\pi i \frac{s}{r}}$, we may introduce another expansion parameter, $\hbar$, with respect to a general root of unity, $e^{2\pi i\frac{s}{r}}$.
That is, we take 
\begin{align}
q = e^{\hbar/r} e^{2\pi i \frac{s}{r}} = e^{\hbar/r} \zeta^{s}_{r}	\,	,
\end{align}
which goes to $\zeta^s_r$ upon $\hbar \rightarrow 0$.
So, in this limit, $q^r \rightarrow 1$.
This was considered in \cite{Chung-rou}, in \cite{Dimofte:2015kkp, Gukov-Pei-rou} in a similar but a slightly different setup, and in \cite{Garoufalidis:2018qds} in the context of Nahm sum.

%%%%%%%%%%%%%%%%%%%%%%%%%%%%%%%%%%%%%%%%%%%%%%%%%%%%%%%%%%%%%%%%%%%%%%%%%%%%%%%%%%%%%%
%%%%%%%%%%%%%%%%%%%%%%%%%%%%%%%%%%%%%%%%%%%%%%%%%%%%%%%%%%%%%%%%%%%%%%%%%%%%%%%%%%%%%%
%%%%%%%%%%%%%%%%%%%%%%%%%%%%%%%%%%%%%%%%%%%%%%%%%%%%%%%%%%%%%%%%%%%%%%%%%%%%%%%%%%%%%%

\subsection{Asymptotic expansion of $q$-Pochhammer symbol at the root of unity}

We note that many knot or link invariants can be expressed in terms of $q$-Pochhammer symbols and monomials.
Examples include trefoil knot $\mathbf{3}_1$, figure-eight knot $\mathbf{4}_1$, $\mathbf{5}_{1,2}$, $\mathbf{6}_{1}$, some other twist knots, and the Hopf links \cite{FGS1,FGS2,FGSS, Nawata:2012pg,GNSSS}. 
Therefore, we study the behavior of the $q$-Pochhammer symbol in the limit that $q$ goes to roots of unity.

%%%%%%%%%%%%%%%%%%%%%%%%%%%%%%%%%%%%%%%%%%%%%%%%%%%%%%%%%%%%%%%%%%%%%%%%%%%%%%%%%%%%%%

\subsubsection*{Asymptotic expansion of $(x;q)_\infty$ around $q \rightarrow 1$}

Before discussing the case of general roots of unity, we review the case that $q \rightarrow 1$.
We denote the deviation of $q$ from 1 as $e^{\epsilon}$, which is $e^{\frac{2\pi i}{K} }$ in this case.

It is known \cite{Zagier-dilogarithm} that the dilogarithm function and the function $\text{Li}_2(x;q)$ are related as
\begin{align}
\lim_{\varepsilon \rightarrow 0} (\varepsilon \text{Li}_2(x;e^{-\varepsilon})) = \text{Li}_2(x)
\end{align}
where
\begin{align}
\text{Li}_{2}(x;q) = \sum_{n=1}^{\infty} \frac{x^n}{n(1-q^n)}		\label{qLi2}
\end{align}
with $x, q \in \mathbb{C}$ with $|x|, |q|<1$.
Also, it can be shown \cite{Zagier-dilogarithm} that 
\begin{align}
-\log(x;q)_\infty = \text{Li}_2(x;q)	\,	.	\label{Poch-Dil}
\end{align}
Therefore, from the generating function of Bernoulli polynomial $B_m(x)$
\begin{align}
\frac{e^{\beta t}}{1-e^t} = - \sum_{m=0}^{\infty} B_m(\beta) \frac{t^{m-1}}{m!}	\label{Bernoulli}
\end{align}
the asymptotic expansion of $\text{Li}_2(q^c x;e^{\epsilon})$ as $\epsilon \rightarrow 0$ is given by
\begin{align}
\text{Li}_2(q^c x;e^{\epsilon}) 
\simeq
-\sum_{m=0}^{\infty} \frac{B_m(c)}{m!} \text{Li}_{2-m}(x) \epsilon^{m-1}	\,	,
\end{align}
so 
\begin{align}
(q^c x;q)_{\infty} \sim \exp \bigg( \sum_{m=0}^{\infty} \frac{B_m(c)}{m!} \text{Li}_{2-m}(x) \epsilon^{m-1} \bigg)	\,	.	\label{Poch-exp}
\end{align}
For example, \eqref{Poch-exp} up to $\epsilon^3$ is given by
\begin{align}
\begin{split}
(q^c x;q)_{\infty} \sim &\exp\Big( \frac{1}{\epsilon} \text{Li}_2(x) + \Big( c - \frac{1}{2} \Big) \log \frac{1}{1-x} + \frac{\epsilon}{12}(6c^2 - 6c + 1) \frac{x}{1-x}  	\\
&+ \frac{\epsilon^2}{12} (2c^3-3c^2+c) \frac{x}{(1-x)^2} + \frac{\epsilon^3}{720}(30c^4-60c^3+30c^2-1) \frac{x+x^2}{(1-x)^3} + \cdots \Big)	\,	.
\end{split}
\end{align}

%%%%%%%%%%%%%%%%%%%%%%%%%%%%%%%%%%%%%%%%%%%%%%%%%%%%%%%%%%%%%%%%%%%%%%%%%%%%%%%%%%%%%%

\subsubsection*{Asymptotic expansion of $(x;q)_\infty$ around roots of unity}

Asymptotic expansion of $(qx;q)_\infty$ has been discussed in \cite{Dimofte:2015kkp, Closset:2018ghr, Chung-rou, Garoufalidis:2018qds}, but has not been applied to the knot polynomials yet.
We express $(q^c x ;q)_\infty$ as
\begin{align}
(q^c x ;q)_\infty = \prod_{a=0}^{r-1} (q^{c+a}x ; q^r)_\infty = \exp \Big( - \sum_{a=0}^{r-1} \text{Li}_2 (q^{c+a} x ; q^r)  \Big)	\,	.	\label{der1}
\end{align}
By using \eqref{qLi2}, we have
\begin{align}
- \sum_{a=0}^{r-1} \text{Li}_2 (q^{c+a} x ; q^r) = -\sum_{a=0}^{r-1} \sum_{n=1}^{\infty} \frac{x^n}{n} \frac{q^{(c+a)n}}{1-q^{rn}}	\,	.	\label{der2}
\end{align}
Since $q^r = e^\hbar$, \eqref{der2} becomes
\begin{align}
-\sum_{a=0}^{r-1}\sum_{n=1}^{\infty} \frac{(\zeta_k^{s(c+a)} x)^n }{n} \frac{e^{n\hbar(c+a)/r}}{1-e^{n\hbar}} 
= \sum_{a=0}^{r-1}\sum_{n=1}^{\infty} \sum_{m=0}^{\infty} \frac{(\zeta_r^{s(c+a)} x)^n }{n^{2-m}} B_m \Big( \frac{c+a}{r} \Big) \frac{\hbar^{m-1}}{m!}	\label{der3}
\end{align}
where \eqref{Bernoulli} is also used.
From $B_m(x+y) = \sum_{l=0}^m \binom{m}{l} B_l(x) y^{m-l}$ where $\binom{m}{l}$ is the binomial coefficient, \eqref{der3} becomes
\begin{align}
\sum_{a=0}^{r-1}\sum_{n=1}^{\infty} \sum_{m=0}^{\infty} \frac{(\zeta_r^{s(c+a)} x)^n }{n^{2-m}} \sum_{l=0}^m \binom{m}{l} B_l(c/r) \left( \frac{a}{r}\right)^{m-l} \frac{\hbar^{m-1}}{m!}	\,	.	\label{der4}
\end{align}
By using the identity $\text{Li}_n(x^k) = k^{n-1} \sum_{j=0}^{k-1} \text{Li}_n(e^{2\pi i \frac{j}{k}} x)$ (\textit{e.g.} in \cite{Lewin}), from \eqref{der4} we obtain
\begin{align}
\sum_{m=0}^{\infty} \frac{B_m(c/r)}{m!} \text{Li}_{2-m}(x^r) r^{m-1} \hbar^{m-1}
+\sum_{m=1}^{\infty} \sum_{l=0}^{m-1} \sum_{a=0}^{r-1} \text{Li}_{2-m}(\zeta_r^{s(c+a)} x) \binom{m}{l} B_l(c/r) \left( \frac{a}{r}\right)^{m-l} \frac{\hbar^{m-1}}{m!}
\end{align}
Therefore, asymptotic expansion of $(q^c x ; q)_\infty$ around $\zeta_{r}^s$ is given by
\begin{align}
\begin{split}
(q^c x ; q)_\infty \sim \exp \Bigg( &\sum_{m=0}^{\infty} \frac{B_m(c/r)}{m!} \text{Li}_{2-m}(x^r) r^{m-1} \hbar^{m-1}	\\
&+\sum_{m=1}^{\infty} \sum_{a=0}^{r-1} \sum_{l=0}^{m-1}  \binom{m}{l} B_l(c/r) \text{Li}_{2-m}(\zeta_r^{s(c+a)} x)  \left( \frac{a}{r}\right)^{m-l} \frac{\hbar^{m-1}}{m!} \Bigg)
\end{split}	\label{q-exp-rou}
\end{align}
For example, up to $\mathcal{O}(\hbar^2)$, \eqref{q-exp-rou} is given by
\begin{align}
\begin{split}
(q^c x ; q)_\infty \sim \exp &\Bigg( 
\frac{1}{\hbar r} \text{Li}_2(x^r) 
-\Big( \frac{c}{r} -\frac{1}{2} \Big) \log (1-x^r) + \sum_{a=0}^{r-1} \frac{a}{r} \log(1-\zeta_r^{s(c+a)} x)	\\
&+\frac{1}{2} \Big( \Big(\frac{c}{r}\Big)^2 - \Big(\frac{c}{r}\Big)+ \frac{1}{2} \Big) \frac{x^r}{1-x^r} r \hbar 
+\sum_{a=0}^{r-1} \frac{1}{2} \frac{a}{r} \frac{\zeta_r^{s(c+a)} x}{1- \zeta_r^{s(c+a)}x} \Big( \frac{a}{r} + 2\Big( \frac{a}{r} -\frac{1}{2} \Big) \Big) \hbar
+\mathcal{O}(\hbar^2)
\Bigg)	
\end{split}
\end{align}
where we see that the leading order of the expansion \eqref{q-exp-rou} is
\begin{align}
\exp \bigg( \frac{1}{\hbar r} \text{Li}_2(x^r) \bigg)	\,	.	\label{leading-exp}
\end{align}

%%%%%%%%%%%%%%%%%%%%%%%%%%%%%%%%%%%%%%%%%%%%%%%%%%%%%%%%%%%%%%%%%%%%%%%%%%%%%%%%%%%%%%

\subsection{Examples: superpolynomial of knots}

We consider the asymptotic expansion discussed above for the superpolynomials of unknot, trefoil knot, and figure-eight knot in the totally symmetric representation $S^n$.
Upon specialization, for example, $a=q^N$, we can recover the Poincar\'e polynomial of a knot for $G=SU(N)$.
There is a subtlety in such specialization \cite{Dunfield:2005si,Gukov:2011ry,FGS1} but there is no problem for the examples here.
We can also take $t=-1$ limit, then we obtain the HOMFLY polynomial or the Jones polynomial upon $a=q^2$.

With the limit $\hbar \rightarrow 0$ in $q = e^{\hbar/r}\zeta_{r}^{s}$, we also need to consider the limit for $x=q^n$ in superpolynomials with the totally symmetric representation $S^n$.
Since we deformed $q$ from $\zeta_r^s$ by $e^{\hbar/r}$, we may consider $x=q^n = e^{\hbar n/r} e^{2\pi i n \frac{s}{r}}$ and take $\hbar \rightarrow 0$ and $n \rightarrow \infty$ such that $\hbar n = u$ is fixed.
We may regard $e^{2\pi i n \frac{s}{r}}$ as still a root of unity in the limit $n \rightarrow \infty$, but this would not be a natural way to consider the limit.
Rather, we may deform $x$ to $x=e^{\frac{\hbar}{r} (n+2\pi i s \xi)}$ where as $\hbar \rightarrow 0$ and $n \rightarrow \infty$, another parameter $\xi$ is taken to $\xi \rightarrow \infty$ such that $\hbar \xi = m \in \mathbb{Z}$ is fixed while $\hbar n = u$ is fixed.
When getting back to the undeformed and original $x$, \textit{i.e.} $x=q^n=\zeta^{ns}_r$, we can take $\hbar \rightarrow 0$ and $\xi \rightarrow \infty$ such that $\hbar \xi = n$ is finite and fixed.
Thus, we take
\begin{align}
x=e^{\hbar n/r} e^{2\pi i \hbar \xi \frac{s}{r}}	\rightarrow 	e^{u/r} e^{2 \pi i \frac{sm}{r}}	\quad	\text{as }	\hbar \rightarrow 0	\,	,	\,	n \rightarrow \infty	\,	,	\,	\xi \rightarrow \infty
\end{align}
such that $\hbar n = u$ is fixed and $\hbar \xi = m \in \mathbb{Z}$ is also fixed.
We also note that in this limit $x^r \rightarrow e^{\hbar n}=e^u$.

With the setup above and also \eqref{q-exp-rou}, we calculate the leading order term of the perturbative expansion of superpolynomials around the root of unity $e^{2\pi i \frac{s}{r}}$.

\subsubsection*{Unknot}

The unnormalized superpolynomial of unknot is 
\begin{align}
\overline{\mathcal{P}}_{0_1}^{S^{n}}(a,q,t) =  a^{-\frac{n}{2}} q^{\frac{n}{2}} (-t)^{-\frac{3}{2}{n}} \frac{(a(-t)^{3};q)_{n}}{(q;q)_{n}}	\,	.
\end{align}
The leading term of asymptotic expansion is given by
\begin{align}
\begin{split}
\overline{\mathcal{P}}_{0_1}^{S^{n}}(a,q,t) \sim \exp \frac{1}{\hbar r} &\bigg( -\frac{1}{2} \log x^r \log a^r - \frac{3}{2} \log x^r \log (-t)^r	\\
&+ \text{Li}_2( a^r (-t)^{3r}) - \text{Li}_2 (a^r (-t)^{3r} x^r) + \text{Li}_2(x^r) - \frac{\pi^2}{6} + \mathcal{O}(\hbar^0) \bigg)
\end{split}
\end{align}
Since $\hbar/r$ of $e^{\hbar/r}$ is the expansion parameter in this setup,
the asymptotic expansion is expressed as
\begin{align}
\mathcal{P}_{\mathcal{K}}^{S^r}(a,q,t) \sim e^{\frac{1}{\hbar/r} \widetilde{\mathcal{W}}}
\end{align}
where the volume function or the twisted superpotential is given by
\begin{align}
\widetilde{\mathcal{W}}(a,x,t) = \int^x_* \log y \frac{dx'}{x'}
\end{align} 
Then, from 
\begin{align}
y= e^{x \frac{\widetilde{\partial \mathcal{W}}}{ \partial x}}
\end{align}
we have
\begin{align}
y^r = a^{-\frac{r}{2}} (-t)^{-\frac{3}{2}r} \frac{1-a^r (-t)^{3r} x^r}{1-x^r} 	\,	.
\end{align}
This is the same with the super A-polynomial of unknot 
\begin{align}
y = a^{-\frac{1}{2}} (-t)^{-\frac{3}{2}} \frac{1-a (-t)^{3} x}{1-x} 	\,	.
\end{align}
with $x\rightarrow x^r$, $y\rightarrow y^r$, $a\rightarrow a^r$, and $-t \rightarrow (-t)^r$.

%%%%%%%%%%%%%%%%%%%%%%%%%%%%%%%%%%%%%%%%%%%%%%%%%%%%%%%%%%%%%%%%%%%%%%%%%%%%%%%%%%%%

\subsubsection*{Trefoil knot}

The case of trefoil knot can be done similarly.
The superpolynomial for the trefoil knot is given by
\begin{align}
\overline{\mathcal{P}}^{\mathcal{S}^{n}}_{3_1}(a,q,t) = \sum _{k=0}^r \frac{(a(-t)^{3};q)_{n}(-a q^{-1} t;q )_{k}}{(q;q)_{k}(q;q)_{n-k}}  a^\frac{n}{2} q^{-\frac{n}{2}}  q^{(r+1) k} (-t)^{2 k -\frac{3n}{2}}
\end{align}
and from the asymptotic expansion we have
\begin{align}
\begin{split}
\widetilde{\mathcal{W}} = \frac{1}{r^2} &\bigg( \log x^r \log a^r + \log z^r \log x^r -\frac{3}{2} \log x^r \log (-t)^r + \log (-t)^{2r} \log z^r -\frac{\pi^2}{6}	\\
&+ \text{Li}_2 (z^r) + \text{Li}_2 (x^r z^{-r}) - \text{Li}_2 (a^r (-t)^{3r} x^r) + \text{Li}_2 (a^r (-t)^{3r}) - \text{Li}_2 (a^r z^r (-t)^r) + \text{Li}_2 (a^r (-t)^r)  \bigg)
\end{split}
\end{align}
From the condition
\begin{align}
1 = e^{z \frac{\partial \widetilde{\mathcal{W}} (a,t, x; z)}{\partial z}}	\,	,	\hspace{5mm}	y = e^{x \frac{\partial \widetilde{\mathcal{W}} (a,t,x;z(a,t,x)) }{\partial x}}	\,	,
\end{align}
we have
\begin{align}
1 = \frac{ x^r (-t)^{2r} (1-x^r z^{-r}) (1- a^r (-t)^r z^r)}{(1-z^r)}	\,	,	\hspace{5mm}	y^r = \frac{a^{\frac{r}{2}} z^{r} (1-a^r (-t)^{3r} x^r ) }{(-t)^{\frac{3r}{2}}(1-x^rz^{-r})}	\,	.
\end{align}
Solving these, we obtain the super A-polynomial
\begin{align}
\begin{split}
0 =& a^{3r/2} (-t)^{5r/2} x^{3r} \left(1-a^{r} (-t)^{3r} x^{r} \right)	\\
&+a^{r} \left(1-(-t)^{2r} x^{r} + 2 (-t)^{2r} \left(1- a^{r} (-t)^{r} \right) x^{2r}-a^{r} (-t)^{5r} x^{3r}+a^{2r} (-t)^{6r} x^{4r} \right) y^{r}	\\
& + a^{r/2} (-t)^{3r/2} (-1+x^{r}) y^{2r}	\,	.
\end{split}
\end{align}
This is the standard super A-polynomial with all the variables are replaced with $r$-th power of them.

%%%%%%%%%%%%%%%%%%%%%%%%%%%%%%%%%%%%%%%%%%%%%%%%%%%%%%%%%%%%%%%%%%%%%%%%%%%%%%%%%%%%

\subsubsection*{Figure-eight knot}

Similarly, from the superpolynomial,
\begin{align}
\overline{\mathcal{P}}_{4_{1}}^{S^{n}}(a,q,t) = \sum_{k=0}^{n} \frac{(a(-t)^{3};q)_{n}}{(q;q)_{k}(q;q)_{n-k}} (aq^{-1}(-t);q)_{k} (aq^{n}(-t)^{3};q)_{k} a^{-k} a^{-\frac{n}{2}} q^{\frac{n}{2}} q^{k(1-n)} (-t)^{-2k} (-t)^{-\frac{3}{2}n}    ,
\end{align}
we obtain
\begin{align}
\begin{split}
\widetilde{\mathcal{W}} = \frac{1}{r^2} \bigg( & -\log z^r \log a^r - \frac{1}{2} \log x^r \log a^r - \log z^r \log x^r - \log z^r \log (-t)^{2r} -  \log x^{r} \log (-t)^{3r/2}	\\
&+ \text{Li}_2 (a^r (-t)^{3r}) - \text{Li}_2 (a^r (-t)^r z^r) + \text{Li}_2 (a^r (-t)^r)	\\
&\hspace{5mm}- \text{Li}_2 (a^r x^r (-t)^{3r} z^r) + \text{Li}_2(z^r) + \text{Li}_2(x^r z^{-r}) - \frac{\pi^2}{3}
 \bigg)	\,	.
\end{split}
\end{align}
This gives
\begin{align}
1=\frac{\left(1-x^{r} z^{-r}\right) (1-a^{r} (-t)^{r} z^{r}) \left(1-a^{r} (-t)^{3r} x^r z^r \right)}{a^r (-t)^{2r} x^r (1-z^r)}	\,	,	\hspace{5mm}		
y^r=\frac{ \left(1-a^r (-t)^{3r} x^r z^r \right)}{a^{r/2} (-t)^{3r/2} z^{r} \left(1-x^r z^{-r}\right) }
\end{align}
and the A-polynomial is given by
\begin{align}
\begin{split}
0 =& a^{3r/2} (-t)^{5r/2} x^{2r} (1-a^r (-t)^{3r} x^r )	\\
&+a^r (-1+(-t)^r (1+(-t)^r) x^r+2 a^r (-t)^{3r} (1-(-t)^r) x^{2r}	\\
&\hspace{10mm}+2 a^r (-t)^{4r} (1-(-t)^r) x^{3r}-a^{2r} (-t)^{6r} (1+(-t)^r) x^{4r}+a^{2r} (-t)^{8r} x^{5r} ) y^r	\\
&+ (1-a^r (-t)^r (1+(-t)^r) x^r+2 a^r (-t)^{2r} (1-(-t)^r) x^{2r}+2 a^{2r} (-t)^{4r} (1-(-t)^r) x^{3r}	\\
&\hspace{10mm}+a^{2r} (-t)^{5r} (1+(-t)^r) x^{4r}-a^{3r} (-t)^{7r} x^{5r} ) y^{2r}
+a^{2r} (-t)^{4r} (-1+x^r) x^{2r} y^{3r}	\,	,
\end{split}
\end{align}
which is the same with the standard super-A-polynomial with variables replaced by their $r$-th powers as expected.

%%%%%%%%%%%%%%%%%%%%%%%%%%%%%%%%%%%%%%%%%%%%%%%%%%%%%%%%%%%%%%%%%%%%%%%%%%%%%%%%%%%%

\subsubsection*{Remarks}

From the examples above, we see that the volume function or the twisted superpotential is given by the standard twisted superpotential up to the overall $1/r^2$ factor where variables are replaced with $r$-th power of them, and also the A-polynomial takes the same form, which could already be expected from \eqref{leading-exp}.
If we want higher order terms in the expansion, we can use \eqref{q-exp-rou} for $q$-Pochhammer symbols.

From the above calculations, we expect in general that in the limit
\begin{align}
q = e^{\hbar/r} \zeta_r^s	\rightarrow	\zeta_r^s	\,	,		\quad	
x=e^{\hbar n/r} e^{2\pi i \hbar \xi \frac{s}{r}}	\rightarrow 	e^{u/r} e^{2 \pi i \frac{sm}{r}}	\label{qx-limit}	
\end{align}
as $\hbar \rightarrow 0$, $n \rightarrow \infty$, and $\xi \rightarrow \infty$ such that $\hbar n = u$ is fixed and $\hbar \xi = m \in \mathbb{Z}$ is also fixed, the asymptotic behavior of the knot invariants would take a form
\begin{align}
P_n(a,q,t) \simeq \exp \bigg( \frac{1}{\hbar r} S_{-1}(u, a^r, (-t)^r) + \sum_{k=0}^{\infty} S_{k}(u,a,t; r, \zeta_r^s, m) \hbar^k \bigg)	\,	.
\end{align}

In the context of the 3d-3d correspondence, the $q$-Pochhammer symbol appears in $S^1 \times D^2$ partition function \cite{Gadde:2013wq, Yoshida:2014ssa, Dimofte:2017tpi}.
The $u$ parameter that appears in the limit of $x$ in \eqref{qx-limit} is proportional to mass parameter of 3d $\mathcal{N}=2$ theory.
Also, since the A-polynomial is a function of $x^r$, which is $x^r=e^u$, the A-polynomial discussed in this paper actually agrees with the A-polynomial obtained in the standard limit $q=e^\epsilon \rightarrow 1$ where $\epsilon n$ is fixed to $u$.
Therefore, matter contents of the theory that are captured in the A-polynomial are the same in either limits.

The physical meaning of the asymptotic expansion around the root of unity and its relation to the corresponding 3d $\mathcal{N}=2$ theory is an interesting problem and we leave it as future work.

%%%%%%%%%%%%%%%%%%%%%%%%%%%%%%%%%%%%%%%%%%%%%%%%%%%%%%%%%%%%%%%%%%%%%%%%%%%%%%%%%%%%
%%%%%%%%%%%%%%%%%%%%%%%%%%%%%%%%%%%%%%%%%%%%%%%%%%%%%%%%%%%%%%%%%%%%%%%%%%%%%%%%%%%%
%%%%%%%%%%%%%%%%%%%%%%%%%%%%%%%%%%%%%%%%%%%%%%%%%%%%%%%%%%%%%%%%%%%%%%%%%%%%%%%%%%%%
%%%%%%%%%%%%%%%%%%%%%%%%%%%%%%%%%%%%%%%%%%%%%%%%%%%%%%%%%%%%%%%%%%%%%%%%%%%%%%%%%%%%
%%%%%%%%%%%%%%%%%%%%%%%%%%%%%%%%%%%%%%%%%%%%%%%%%%%%%%%%%%%%%%%%%%%%%%%%%%%%%%%%%%%%

\section{Discussion}
\label{discussion}

We discussed the WRT invariant for an infinite family of Seifert manifolds at other roots of unity in terms of homological blocks.
The structure is the same but with a few differences. 
One is obviously the limit $q\searrow e^{2\pi i \frac{s}{r}}$, another is about the factor $\sum_{v=0}^{s-1} e^{2\pi i \frac{r}{s} \frac{P}{H} (Hv+u)^2}$.
The properties in the case of the standard root of unity, such as the symmetries from the action of the center and the complex conjugation, also hold.

In the second half, we discussed an asymptotic expansion of knot invariants around general roots of unity where the limit we took is different from the limit in the standard volume conjectures.
We calculated the leading order of the asymptotic expansions of superpolynomials around the roots of unity $e^{2\pi i \frac{s}{r}}$ and saw that the A-polynomials are the same as those in the context of standard volume conjecture upon replacement of variables by their $r$-th powers and similarly for the volume function/twisted superpotential up to an overall factor.	\\

There are several interesting directions.
It would be interesting to generalize the $SU(2)$ case to higher rank cases and also to the case of the two-variable series for knot complements in \cite{Gukov:2019mnk}.
Detailed resurgent analysis for the case of rational $K$ would also be interesting.
In addition, the physical understanding for the case of rational $K$ via the 3d-3d correspondence would be an important direction to study.

More thorough study on the asymptotic expansion discussed in section \ref{asymp-exp} would be interesting.
For example, systematic calculations of higher order terms in the asymptotic expansion around roots of unity by using quantization of A-polynomial would be an interesting direction.
It would be important to study mathematical meaning or applications of such expansion, and also physical meaning of the expansion in the context of the 3d-3d correspondence.

%%%%%%%%%%%%%%%%%%%%%%%%%%%%%%%%%%%%%%%%%%%%%%%%%%%%%%%%%%%%%%%%%%%%%%%%%%%%%%%%%%%%
%%%%%%%%%%%%%%%%%%%%%%%%%%%%%%%%%%%%%%%%%%%%%%%%%%%%%%%%%%%%%%%%%%%%%%%%%%%%%%%%%%%%
%%%%%%%%%%%%%%%%%%%%%%%%%%%%%%%%%%%%%%%%%%%%%%%%%%%%%%%%%%%%%%%%%%%%%%%%%%%%%%%%%%%%
%%%%%%%%%%%%%%%%%%%%%%%%%%%%%%%%%%%%%%%%%%%%%%%%%%%%%%%%%%%%%%%%%%%%%%%%%%%%%%%%%%%%
%%%%%%%%%%%%%%%%%%%%%%%%%%%%%%%%%%%%%%%%%%%%%%%%%%%%%%%%%%%%%%%%%%%%%%%%%%%%%%%%%%%%

\acknowledgments{I would like to thank Sergei Gukov and Du Pei for valuable discussion.
I am also grateful to the Korea Institute for Advanced Study (KIAS) and Centre for Quantum Geometry of Moduli Spaces (QGM) at Aarhus University for hospitality at some stages of this work.
}

\bibliographystyle{JHEP}
\bibliography{ref}

\providecommand{\href}[2]{#2}\begingroup\raggedright\begin{thebibliography}{10}

\bibitem{Witten-Jones}
E.~Witten, \emph{Quantum field theory and the jones polynomial}, {\emph{Comm.
  Math. Phys.} {\bfseries 121} (1989) 351--399}.

\bibitem{RT}
N.~Reshetikhin and V.~G. Turaev, \emph{Invariants of {$3$}-manifolds via link
  polynomials and quantum groups},
  \href{https://doi.org/10.1007/BF01239527}{\emph{Invent. Math.} {\bfseries
  103} (1991) 547--597}.

\bibitem{Gukov-Putrov-Vafa}
S.~Gukov, P.~Putrov and C.~Vafa, \emph{{Fivebranes and 3-manifold homology}},
  \href{https://doi.org/10.1007/JHEP07(2017)071}{\emph{JHEP} {\bfseries 07}
  (2017) 071}, [\href{https://arxiv.org/abs/1602.05302}{{\ttfamily
  1602.05302}}].

\bibitem{Gukov-Pei-Putrov-Vafa}
S.~Gukov, D.~Pei, P.~Putrov and C.~Vafa, \emph{{BPS spectra and 3-manifold
  invariants}},  \href{https://arxiv.org/abs/1701.06567}{{\ttfamily
  1701.06567}}.

\bibitem{Gukov-Marino-Putrov}
S.~Gukov, M.~Marino and P.~Putrov, \emph{{Resurgence in complex Chern-Simons
  theory}},  \href{https://arxiv.org/abs/1605.07615}{{\ttfamily 1605.07615}}.

\bibitem{Chun:2017dbf}
S.~Chun, \emph{{A resurgence analysis of the $SU(2)$ Chern-Simons partition
  functions on a Brieskorn homology sphere $\Sigma(2,5,7)$}},
  \href{https://arxiv.org/abs/1701.03528}{{\ttfamily 1701.03528}}.

\bibitem{Cheng:2018vpl}
M.~C.~N. Cheng, S.~Chun, F.~Ferrari, S.~Gukov and S.~M. Harrison, \emph{{3d
  Modularity}},  \href{https://arxiv.org/abs/1809.10148}{{\ttfamily
  1809.10148}}.

\bibitem{Chung-wrt}
H.-J. Chung, \emph{{BPS Invariants for Seifert Manifolds}},
  \href{https://arxiv.org/abs/1811.08863}{{\ttfamily 1811.08863}}.

\bibitem{Gukov:2019mnk}
S.~Gukov and C.~Manolescu, \emph{{A two-variable series for knot complements}},
   \href{https://arxiv.org/abs/1904.06057}{{\ttfamily 1904.06057}}.

\bibitem{Lawrence-Rozansky}
R.~Lawrence and L.~Rozansky, \emph{Witten-reshetikhin-turaev invariants of
  seifert manifolds}, \href{https://doi.org/10.1007/s002200050678}{\emph{Comm.
  Math. Phys.} {\bfseries 205} (1999) 287--314}.

\bibitem{Chung-rou}
H.-J. Chung, \emph{{Unpublished note, 2018}}, .

\bibitem{Gukov-Pei-rou}
S.~Gukov and D.~Pei, \emph{{Unpublished note, 2017}}, .

\bibitem{Dimofte:2015kkp}
T.~Dimofte and S.~Garoufalidis, \emph{{Quantum modularity and complex
  Chern–Simons theory}},
  \href{https://doi.org/10.4310/CNTP.2018.v12.n1.a1}{\emph{Commun. Num. Theor.
  Phys.} {\bfseries 12} (2018) 1--52},
  [\href{https://arxiv.org/abs/1511.05628}{{\ttfamily 1511.05628}}].

\bibitem{Garoufalidis:2018qds}
S.~Garoufalidis and D.~Zagier, \emph{{Asymptotics of Nahm sums at roots of
  unity}},  \href{https://arxiv.org/abs/1812.07690}{{\ttfamily 1812.07690}}.

\bibitem{Kucharski:2019fgh}
P.~Kucharski, \emph{{$\hat{Z}$ invariants at rational $\tau$}},
  \href{https://arxiv.org/abs/1906.09768}{{\ttfamily 1906.09768}}.

\bibitem{Lawrence-Zagier}
R.~Lawrence and D.~Zagier, \emph{Modular forms and quantum invariants of
  {$3$}-manifolds},
  \href{https://doi.org/10.4310/AJM.1999.v3.n1.a5}{\emph{Asian J. Math.}
  {\bfseries 3} (1999) 93--107}.

\bibitem{Zagier-identity}
D.~Zagier, \emph{Vassiliev invariants and a strange identity related to the
  {D}edekind eta-function},
  \href{https://doi.org/10.1016/S0040-9383(00)00005-7}{\emph{Topology}
  {\bfseries 40} (2001) 945--960}.

\bibitem{Hikami-Kirilov}
K.~Hikami and A.~N. Kirillov, \emph{Torus knot and minimal model},
  \href{https://doi.org/https://doi.org/10.1016/j.physletb.2003.09.007}{\emph{Physics
  Letters B} {\bfseries 575} (2003) 343 -- 348}.

\bibitem{Hikami-torus}
K.~Hikami, \emph{Quantum invariant for torus link and modular forms},
  \href{https://doi.org/10.1007/s00220-004-1046-2}{\emph{Comm. Math. Phys.}
  {\bfseries 246} (2004) 403--426}.

\bibitem{Hikami-lattice1}
K.~Hikami, \emph{Quantum invariant, modular form, and lattice points},
  \href{https://doi.org/10.1155/IMRN.2005.121}{\emph{Int. Math. Res. Not.}
  (2005) 121--154}.

\bibitem{Hikami-lattice2}
K.~Hikami, \emph{Quantum invariants, modular forms, and lattice points. ii},
  \href{https://doi.org/10.1063/1.2349484}{\emph{J. Math. Phys.} {\bfseries 47}
  (2006) 102301, 32}.

\bibitem{DG-review}
T.~Dimofte and S.~Gukov, \emph{{Quantum Field Theory and the Volume
  Conjecture}}, {\emph{Contemp. Math.} {\bfseries 541} (2011) 41--67},
  [\href{https://arxiv.org/abs/1003.4808}{{\ttfamily 1003.4808}}].

\bibitem{FGS1}
H.~Awata, S.~Gukov, P.~Sulkowski and H.~Fuji, \emph{{Volume Conjecture: Refined
  and Categorified}},
  \href{https://doi.org/10.4310/ATMP.2012.v16.n6.a3}{\emph{Adv. Theor. Math.
  Phys.} {\bfseries 16} (2012) 1669--1777},
  [\href{https://arxiv.org/abs/1203.2182}{{\ttfamily 1203.2182}}].

\bibitem{FGS2}
H.~Fuji, S.~Gukov and P.~Sulkowski, \emph{{Super-A-polynomial for knots and BPS
  states}}, \href{https://doi.org/10.1016/j.nuclphysb.2012.10.005}{\emph{Nucl.
  Phys.} {\bfseries B867} (2013) 506--546},
  [\href{https://arxiv.org/abs/1205.1515}{{\ttfamily 1205.1515}}].

\bibitem{FGSS}
H.~Fuji, S.~Gukov, M.~Stosic and P.~Sulkowski, \emph{{3d analogs of
  Argyres-Douglas theories and knot homologies}},
  \href{https://doi.org/10.1007/JHEP01(2013)175}{\emph{JHEP} {\bfseries 01}
  (2013) 175}, [\href{https://arxiv.org/abs/1209.1416}{{\ttfamily 1209.1416}}].

\bibitem{Nawata:2012pg}
S.~Nawata, P.~Ramadevi, Zodinmawia and X.~Sun, \emph{{Super-A-polynomials for
  Twist Knots}}, \href{https://doi.org/10.1007/JHEP11(2012)157}{\emph{JHEP}
  {\bfseries 11} (2012) 157},
  [\href{https://arxiv.org/abs/1209.1409}{{\ttfamily 1209.1409}}].

\bibitem{GNSSS}
S.~Gukov, S.~Nawata, I.~Saberi, M.~Stošić and P.~Sułkowski,
  \emph{{Sequencing BPS Spectra}},
  \href{https://doi.org/10.1007/JHEP03(2016)004}{\emph{JHEP} {\bfseries 03}
  (2016) 004}, [\href{https://arxiv.org/abs/1512.07883}{{\ttfamily
  1512.07883}}].

\bibitem{Zagier-dilogarithm}
D.~Zagier, \emph{The dilogarithm function},  in \emph{Frontiers in number
  theory, physics, and geometry. {II}}, pp.~3--65.
\newblock Springer, Berlin, 2007.
\newblock \href{https://doi.org/10.1007/978-3-540-30308-4_1}{DOI}.

\bibitem{Closset:2018ghr}
C.~Closset, H.~Kim and B.~Willett, \emph{{Seifert fibering operators in 3d
  $\mathcal{N}=2$ theories}},
  \href{https://doi.org/10.1007/JHEP11(2018)004}{\emph{JHEP} {\bfseries 11}
  (2018) 004}, [\href{https://arxiv.org/abs/1807.02328}{{\ttfamily
  1807.02328}}].

\bibitem{Lewin}
L.~Lewin, \emph{Polylogarithms and associated functions}.
\newblock North-Holland Publishing Co., New York-Amsterdam, 1981.

\bibitem{Dunfield:2005si}
N.~M. Dunfield, S.~Gukov and J.~Rasmussen, \emph{{The Superpolynomial for knot
  homologies}},  \href{https://arxiv.org/abs/math/0505662}{{\ttfamily
  math/0505662}}.

\bibitem{Gukov:2011ry}
S.~Gukov and M.~Stošić, \emph{{Homological Algebra of Knots and BPS States}},
  \href{https://doi.org/10.1090/pspum/085/1377,
  10.2140/gtm.2012.18.309}{\emph{Proc. Symp. Pure Math.} {\bfseries 85} (2012)
  125--172}, [\href{https://arxiv.org/abs/1112.0030}{{\ttfamily 1112.0030}}].

\bibitem{Gadde:2013wq}
A.~Gadde, S.~Gukov and P.~Putrov, \emph{{Walls, Lines, and Spectral Dualities
  in 3d Gauge Theories}},
  \href{https://doi.org/10.1007/JHEP05(2014)047}{\emph{JHEP} {\bfseries 05}
  (2014) 047}, [\href{https://arxiv.org/abs/1302.0015}{{\ttfamily 1302.0015}}].

\bibitem{Yoshida:2014ssa}
Y.~Yoshida and K.~Sugiyama, \emph{{Localization of 3d $\mathcal{N}=2$
  Supersymmetric Theories on $S^1 \times D^2$}},
  \href{https://arxiv.org/abs/1409.6713}{{\ttfamily 1409.6713}}.

\bibitem{Dimofte:2017tpi}
T.~Dimofte, D.~Gaiotto and N.~M. Paquette, \emph{{Dual boundary conditions in
  3d SCFT\textquoteright{}s}},
  \href{https://doi.org/10.1007/JHEP05(2018)060}{\emph{JHEP} {\bfseries 05}
  (2018) 060}, [\href{https://arxiv.org/abs/1712.07654}{{\ttfamily
  1712.07654}}].

\end{thebibliography}\endgroup

\end{document}